\documentclass[11pt]{article}
\usepackage{amsmath,amssymb,amsfonts}
\usepackage[T1]{fontenc}
\usepackage[latin1]{inputenc}
\usepackage{epsfig,psfrag}
\numberwithin{equation}{section}
\newtheorem{thm}{Theorem}[section]

\newtheorem{alem}[thm]{Lemma}
\newtheorem{aprop}[thm]{Proposition}
\newtheorem{acor}[thm]{Corollary}
\newtheorem{arem}[thm]{Remark}
\newenvironment{adem}[1][]%
   {\ \\ {\bf Proof #1. }}%
   {\hfill\mbox{\rule{2 true mm}{3 true mm}}}
   {\ \\ {\bf Example #1. }}%
   {\hfill\mbox{\rule{2 true mm}{3 true mm}}}
\newcommand{\R}{\mathbb{R}}
\renewcommand{\P}{\mathbb{P}}

\newcommand{\E}{{\mathbb E}}

\newcommand{\N}{{\mathbb N}}

\newcommand{\mycomments}[1]{  }

\newcommand{\be}{ \begin{eqnarray*}  }
\newcommand{\ee}{ \end{eqnarray*}  }
\renewcommand{\S}{{\bar{S}}}
\newcommand{\U}{{\bar{u}}}
\renewcommand{\Bbb}{\mathbb}

 \title{Regularity of the Exercise Boundary for\\ American Put Options on Assets with Discrete Dividends}
\author{B. Jourdain\thanks{Université Paris-Est, CERMICS, Projet MathFi
    ENPC-INRIA-UMLV, 6 et 8 avenue Blaise Pascal, 77455 Marne La Vallée, Cedex
    2, France, e-mail : jourdain@cermics.enpc.fr.  This research benefited
    from the support of the ``Chair Risques Financiers'', Fondation du
    Risque. Research was partially completed while the author was visiting the Institute for Mathematical Sciences, National University of Singapore in 2009. The visit was supported by the institute.
} \ and M. Vellekoop\thanks{Corresponding Author, University of Amsterdam,
Faculty of Economics and Business,
Department of Quantitative Economics,
Section Actuarial Science,
Roetersstraat 11,
1018 WB Amsterdam, e-mail : m.h.vellekoop@uva.nl}}

\begin{document}
\maketitle

\begin{abstract}
We analyze the regularity of the optimal exercise boundary for the American Put option 
when the underlying asset pays a discrete dividend at a known
time $t_d$ during the lifetime of the option. The ex-dividend asset price
process is assumed to follow Black-Scholes dynamics and the dividend
amount is a deterministic function of the ex-dividend asset price just
before the dividend date. The solution to the associated optimal
stopping problem can be characterised in terms of an optimal exercise
boundary which, in contrast to the case when there are no dividends, may
no longer be monotone. In this paper we prove that when the dividend
function is positive and concave, then the boundary is non-increasing in a left-hand
neighbourhood of $t_d$, and tends to
$0$ as time tends to $t_d^-$ with a speed that we can
characterize. When the dividend function is linear in a neighbourhood of zero, then we show continuity of the exercise boundary and a high contact principle in the left-hand neighbourhood of $t_d$. When it is globally linear, then right-continuity of the boundary and the high contact principle are proved to hold globally. Finally, we show how all the previous results can be extended to multiple 
dividend payment dates in that case.
\end{abstract}

\section*{Introduction}

We consider the American Put option with strike $K>0$ and maturity $T>0$
on an underlying stock. We assume that the stochastic dynamics of the ex-dividend price process of this stock can be modelled by
the Black-Scholes model and that at the $I\in {\mathbb N}$ given times $t_d^I<t_d^{I-1}<...<t_d^1$ in the time interval $(0,T)$, discrete stock dividends are paid. 
The case without dividends is denoted by $I=0$ and we will use the convention that $t_d^{I+1}=0$ and $t_d^0=T$ throughout the paper. The value of the dividend payments 
are functions $D^j:\R_+\rightarrow \R_+$ ($1\leq j\leq I$) of the ex-dividend asset price. This means that the stock price process satisfies
\begin{equation} dS_u=\sigma S_u dW_u+rS_udu-\sum_{j=1}^I D^j(S_{u-})d1_{\{u\geq t_d^j\}} \label{eq:S} \end{equation}
for an initial price $S_0$, interest rate $r$ and volatility $\sigma$ which are assumed to be
positive and with $W$ a standard Brownian Motion.

Throughout the paper we assume that 
the dividend functions $D^j$ are non-negative and non-decreasing for all $1\leq j\leq I$
and such that $x\in\R_+\mapsto x-D^j(x)$ is non-negative and non-decreasing.
We will pay particular attention to the following special cases :
\begin{itemize}
   \item $D^j(x)=(1-\rho_j)x$ where $\rho_j\in(0,1)$, which we will call the {\sl proportional}
     dividend case, 
\item $D^j(x)=D^j\wedge x$ with $D^j>0$, which we will call the {\sl constant} dividend case, and
\item $D^j(x)=\min\{a_j+b_jx,c_jx\}$ with $a_j,b_j,c_j\geq 0$ and $c_j\leq 1$, which we call the {\sl mixed} dividend case.
\end{itemize}
We will see that the behaviour of $D^j$ around zero determines the behaviour
 of the exercise boundary at the dividend dates $t_d^j$ so the latter case will turn out to be very similar to the one where we have proportional dividends.

For $t\in[0,T]$, let \begin{equation}\label{defu}
U_t = {\rm ess}.\sup_{\tau \in {\cal T}_{[t,T]}} {\Bbb E}[
e^{-r(\tau-t)}(K-S_\tau)^+|{\cal F}_t]
\end{equation} denote the price at time
$t$ of the American Put option, where ${\cal T}_{[t,T]}$ is the set of stopping times with respect to the
filtration ${\cal F}_t\stackrel{\rm def}{=}\sigma(W_s,0\leq s\leq t)$
taking values in $[t,T]$.
The solution to this optimal stopping problem for the case without
dividends (i.e. $I=0$) goes back to the work of McKean \cite{McKean} and Van Moerbeke
\cite{VanMoerbeke}. The optimal stopping time is the first time that the
asset price process falls below a time-dependent value (the so-called
exercise boundary which we will denote by $c^0$), and McKean
derived a free-boundary problem involving both the pricing function
$u^0$ such that $U_t=u^0(t,S_t)$ and $c^0$. Van Moerbeke
derived an integral equation which involves both $c^0$ and its
derivative, but in later work by Kim \cite{Kim}, Jacka \cite{Jacka} and
Carr, Jarrow and Myneni \cite{CarrJarrowMyneni} an integral equation was
derived which only involves $c^0$ itself. The regularity and
uniqueness of solutions to this equation was left as an open problem in
those papers. Uniqueness was proven by Peskir
\cite{PeskirAmericanOption}, using his change-of-variable formula with
local time on curves \cite{PeskirLocalTimeSpace}. It is known that the
optimal exercise boundary is convex
\cite{ChenChadamJiangZheng,EkstromConvexBoundary} and its asymptotic
behaviour at maturity is given in \cite{Lamberton}. But although it was
claimed in several papers (for example \cite{Myneni}) that it is $C^1$
at all points prior to maturity, a complete proof has been given only
recently by Chen and Chadam \cite{ChenChadam}. In fact, in that paper it
was actually shown that it is $C^\infty$ in all those points and a later
paper by Bayraktar and Xing \cite{BayraktarXing} shows that this remains
true if the underlying asset pays continuous dividends at a fixed
rate.
In practice, continuous dividends are not a satisfying model since
dividends are paid once a year or quarterly. That is why we are
interested in dividends that are paid at a number of discrete points in time. 
To begin with, we deal in this paper
with the simplest situation where there is only one dividend time $t_d^1$
before the maturity $T$ of the Put option\footnote{When there is only one dividend
date, i.e. $I=1$, we will often suppress the value $i=1$ in our notation, 
so we will write $t_d$ instead of $t_d^1$, $D$ for $D^1$ and so on.}. 
Afterwards we show how some results can be extended to the case of multiple dividends.

When we assume discrete dividend payments such as the proportional
 or fixed dividend payments mentioned above, the optimal exercise
 boundary will become discontinuous at the dividend dates and before the
 dividend dates it may not be monotone (see Figure \ref{fig:exbound}).
Integral formulas  for the
 exercise boundary which are similar to the ones in
 \cite{CarrJarrowMyneni} have been derived under the assumption that the
 boundary is Lipschitz continuous (see G\"ottsche and Vellekoop
 \cite{GoettscheVellekoop}) or locally monotonic (Vellekoop \&
 Nieuwenhuis \cite{VellekoopNieuwenhuisGeneralAmPutPremium}). In this
 paper we therefore study conditions under which such regularity
 properties of the optimal exercise boundary under discrete dividend
 payments can be proven.

In the first Section, we introduce the pricing functions $u^i:[0,T]\times {\Bbb R}_+$ of the
American Put option in the model \eqref{eq:S} for $0\leq i\leq I$ and the associated
exercise boundaries $c^i$ where the $i$ means for $i\geq 1$ that only at the times $t_d^i,t_d^{i-1},...,t_d^1$ dividends are being paid while $i=0$ means that no dividends are being paid. 
 We then explain that for $I\geq i\geq 1$, on the time-interval
 $[t_d^{i+1},t_d^{i})$, the American Put
price $u^i$ is equal to the price of an American option in the Black-Scholes model with no
dividends if we take its maturity $t_d^i$ and its payoff $x\mapsto (K-x)^+$ when exercised early and a modified
payoff $x\mapsto {u}^{i-1}(t_d^i,x-D^i(x))$ when
exercised at the maturity time $t_d^i$.  Studying the properties of the single dividend case will then allow us to derive properties of the sequence of functions $u^i$ and $c^i$ in a recursive manner.

In the second Section, we therefore first look at the single dividend case only and prove that when the dividend function is positive and concave, then the boundary is non-increasing in a left-hand
neighbourhood of $t_d$, and tends to
$0$ as time tends to $t_d^-$ with a speed that we can
characterize. In the third Section we assume moreover that the dividend function is linear in a
neighbourhood of $0$, a condition satisfied in the {\sl
  proportional}, the {\sl constant} and the {\sl mixed} dividend cases. Then we show that the exercise boundary is continuous
and a high contact principle holds in a left-hand
neighbourhood of $t_d$. In the {\sl proportional} dividend case, right-continuity of the boundary and the high contact principle are proved to hold globally. Finally, we show how results for a single dividend date can be extended to multiple dividend dates
in that case.
\medskip

\underline{\bf Notations and definitions :}
\begin{itemize}
   \item For $t\in [0,T]$ and $x\geq 0$, we use the notation $\S^x_t=xe^{\sigma
  W_t+(r-\frac{\sigma^2}{2})t}$ for the stock price at time $t$  when
the initial price is equal to $x$ and when there is no
dividend (i.e. $I=0$). We also denote by $L_t^y(\S^x)$ the local time at level $y>0$ and
time $t$ of the process $\S^x$ and by $p(t,y)=\frac{1_{\{y>0\}}}{\sigma y\sqrt{2\pi
    t}}\exp\left(-\frac{(\log(y/x)-(r-\frac{\sigma^2}{2})t)^2}{2\sigma^2
    t}\right)$ the density of $\S^x_t$ with respect to the
Lebesgue measure when $t,x>0$.\item Let ${\cal A}$ denote the infinitesimal generator of the
  Black-Scholes model without dividends : ${\cal
  A}f(x)=\frac{\sigma^2x^2}{2}f''(x)+rxf'(x)-rf(x)$.
\item If $(t,x)\in [0,T]\times\R_+$ and $1\leq i\leq I$ we write  $S^{x,t,i}_u$ for the solution to 
\begin{equation} dS^{x,t,i}_v=\sigma S^{x,t,i}_v dW_v+rS^{x,t,i}_vdv-\sum_{j=1}^i D^j(S^{x,t,i}_{v^-})d1_{\{v\geq t_d^j\}} 
\end{equation}
 for $u\geq t$
 under the initial condition that $S^{x,t,i}_t=x$. Note that we still retain the notation introduced in (\ref{eq:S}) so $S_u=S^{S_0,0,I}_u$.
\item Let $N(y)=\int_{-\infty}^y e^{-z^2/2}\frac{dz}{\sqrt{2\pi}}$ be the
cumulative distribution function of the standard normal law.
\item Let $C$ denote a constant with may change from line to line.
\item We say that $D:\R_+\to\R_+$ is positive when $\forall x>0$, $D(x)>0$.
\item By a left-hand neighbourhood of $x\in\R$, we mean an open interval
$(x-\varepsilon,x)$ for some $\varepsilon>0$.
\item We will often denote the value function $u^0$ for the case without dividends by $\bar{u}$ and the value function $u^1$ for the case of one dividend by $u$.\end{itemize}

\section{Preliminary results}
The following results, which have been proven in
\cite{ElKarouiBook,ElKarouiLepeltierMillet,Hamadene,PeskirShiryaevBook},
provide an optimal stopping time in \eqref{defu}.

\begin{aprop}\label{snell}
Let  $\{G_t,\, t\in [0,T]\}$ be an $({\cal F}_t)$-adapted
right-continuous upper-semicontinuous process with ${\Bbb E}(\sup_{t\in
  [0,T]} |G_t|)<\infty$.\\Then the c\`adl\`ag version of the Snell
envelope $U_t={\rm ess.}\sup_{\tau\in {\cal T}_{[t,T]}}{\Bbb
  E}(G_\tau\mid{\cal F}_t)$ is continuous on $[0,T]$ and the stopping
time $\tau=\inf\{ s\geq t: U_s=G_s\}$ is optimal : $U_t={\Bbb
  E}(G_\tau\mid{\cal F}_t)$. 
\end{aprop}

\begin{figure}[!ht]
\begin{center}
\scalebox{0.75}{\includegraphics{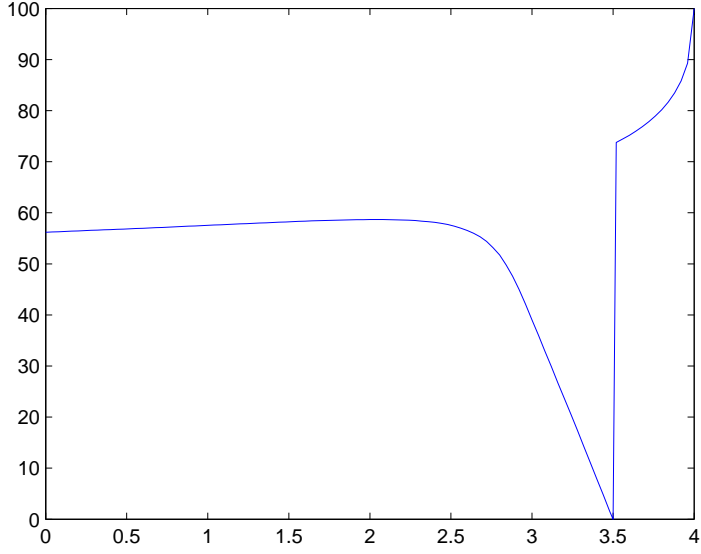}}
\caption{Exercise boundary $t\mapsto {c^1(t)}$ ($K=100$, $T=4$, $t^1_d=3.5$, $r=0.04$,
  $\sigma=0.3$, {\sl proportional} dividends : $\rho_1=0.05$) obtained by a
  binomial tree method\label{fig:exbound}}
\end{center}\end{figure}

 The conditions for this result are satisfied by
 $G_t^i=e^{-rt}(K-S_t^{S_0,0,i})^+$ since for all $t\in [0,T]$ we have $|G_t^i|\leq K$
 and $G_t^i$ is right-continuous and upper semicontinuous for all $t\in
 [0,T]$ since the jump sizes of $S_t^{S_0,0,i}$ at $t=t_d^j$  are non-positive for all $1\leq j\leq i$ (for a Call option $G_t^i=e^{-rt}(S_t^{S_0,0,i}-K)^+$ is no longer upper-semicontinuous and, in the single dividend case, Battauz and Pratelli \cite{BattauzPratelli} artificially stretch
the time-interval $[0, T]$ by introducing a ficticious interval $[\underline{t}^1_d,\bar{t}^1_d]$ where $\underline{t}^1_d$ denotes the end of the cum-dividend date and $\bar{t}^1_d$ the beginning of the ex-dividend date to reduce the evaluation problem to the computation of the Snell envelope on stopping times taking values in $[0,\underline{t}^1_d]\cup[\bar{t}^1_d,T]$).
According to \cite{ElKarouiLepeltierMillet}, there thus exists pricing
functions $u^i$ defined as follows:
\begin{aprop}\label{pricfunc}
   Take $1\leq i\leq I$ and a constant $S_0>0$. The Snell envelop $U^i$ of $\{G_t^i=e^{-rt}(K-S^{S_0,0,i}_t)^+,\, t\in [0,T]\}$ is such
   that $U^i_t=e^{-rt}u(t,S^{S_0,0,i}_t)$ where $$\forall (t,x)\in[0,T]\times\R_+,\;u^i(t,x)\stackrel{\rm
     def}{=}\sup_{\tau\in{\cal
       T}_{[t,T]}}\E(e^{-r(\tau-t)}(K-S^{x,t,i}_\tau)^+).$$
Moreover the previous supremum is attained for $\tau=\inf\{s\geq t:u^i(s,S^{x,t,i}_s)=(K-S^{x,t,i}_s)^+\}$.
\end{aprop}
Let us now derive some properties of the pricing functions $u^i$ and define the exercise boundaries $c^i$.
\begin{alem}\label{defg}
Let for all $1\leq j\leq I$ the dividend functions $D^j$ be non-negative, non-decreasing
and such that $x\in\R_+\mapsto x-D^j(x)$ is non-negative and
non-decreasing. Then we have  
 \begin{equation}
   \forall 0\leq i\leq I,\; \forall
t\in [0,T],\;\forall x>y\geq 0,\;0\leq u^i(t,y)-u^i(t,x)\leq x-y.\label{lip}
\end{equation} 
For $t\in[0,T]$, let \[ c^i(t)=\inf\{x>0:u^i(t,x)>(K-x)^+\}. \] 
Then $c^i(t)<K$ for $t\in[0,T)$ and we have that $\{x\geq 0:u^i(t,x)=(K-x)^+\}=[0,c^i(t)]$. Last the functions $c^i$
cannot vanish on an interval.\end{alem}
Figure \ref{fig:exbound} plots the exercise boundary $t\mapsto c^1(t)$ of
the Put option with strike $K=100$ and maturity $T=4$ in the model
\eqref{eq:S} with $r=0.04$, $\sigma=0.3$, $t^1_d=3.5$ and {\sl proportional} dividends
with $\rho_1=0.05$. This exercise boundary was computed by a binomial tree
method (see \cite{VellekoopNieuwenhuisTree}).

\begin{adem}
For the first part, we use a similar proof as in
\cite{GoettscheVellekoop}. For a fixed $t\in [0,T]$ take $x>y\geq 0$ which,
with the monotonicity of $z\mapsto z-D^j(z)$ for all $1\leq j\leq I$ implies that
$S^{x,t,i}_v\geq S^{y,t,i}_v$ for all $v\in[t,T]$. Now fix the value of $i$ with $0\leq i\leq I$.
For $\tau_x\in {\cal T}_{[t,T]}$ such
that $u^i(t,x)=
{\Bbb E}[ e^{-r(\tau_x-t)}(K-S^{x,t,i}_{\tau_x})^+]$, since $\tau_x$ need not be optimal for the case where the stock price at time $t$ 
equals $y$, we deduce
$$u^i(t,x)-u^i(t,y)\leq{\Bbb E}[
e^{-r(\tau_x-t)}((K-S^{x,t,i}_{\tau_x})^+-(K-S^{y,t,i}_{\tau_x})^+) ]\leq
0.$$  For $\tau_y\in {\cal T}_{[t,T]}$ such that 
$
u^i(t,y)={\Bbb E}[ e^{-r(\tau_y-t)}(K-S^{y,t,i}_{\tau_y})^+]$,\be
u^i(t,y)-u^i(t,x)  &\leq & 
{\Bbb E}[ e^{-r(\tau_y-t)}(K-S^{y,t,i}_{\tau_y})^+]-{\Bbb E}[ e^{-r(\tau_y-t)}(K-S^{x,t,i}_{\tau_y})^+]\\
&\leq &
 {\Bbb E}[ e^{-r(\tau_y-t)} (S_{\tau_y}^{x,t,i}-S_{\tau_y}^{y,t,i})]\\
&=& x-y - \sum_{j=1}^i {\Bbb E}[  e^{-r(\tau_y-t)}1_{\{\tau_y\geq t_d^j>t\}}( D^j(S_{t_d^j-}^{x,t,i})-D^j(S_{t_d^j-}^{y,t,i})) \, \bar{S}_{\tau_y-t_d^j}^{1}]
\ \leq \  x-y
\ee
because $S_{t_d^j-}^{x,t,i}\geq S_{t_d^j-}^{y,t,i}$ and the function $D^j$ is non-decreasing. 

Since $u^i(t,x)\geq (K-x)^+$ for all $t\in [0,T]$ and $x\geq 0$, the
definition of $c^i(t)$ implies that $u^i(t,x)=(K-x)^+$ for $x\in [0,c^i(t))$ and by the continuity of $x\to u^i(t,x)-(K-x)^+$
this must then be true for $x=c^i(t)$ as well when $c^i(t)>0$. When
$c^i(t)=0$, $u^i(t,c^i(t))=K=(K-c^i(t))^+$. If $x>c^i(t)$ then, by
definition of $c^i(t)$ there
exists $y\in (c^i(t),x]$ such that $u^i(t,y)>(K-y)^+$ and $u^i(t,x)\geq
u^i(t,y)+y-x> K-x$. For $t\in[0,T)$, since $u^i(t,x)\geq \E(e^{-r(T-t)}(K-S^{x,t,i}_T)^+)>0$,
one deduces that $u^i(t,x)>(K-x)^+$ for $x>c^i(t)$ and that $c^i(t)<K$. Last, $c^i(T)=+\infty$.\\
Assume that there exists an interval $[t_1,t_2)$ with $0\leq t_1\leq
t_2\leq T$ such
that $c^i$ is zero in every point of this interval, and for $x>0$, let $\tau_x\in {\cal T}_{[t_1,T]}$ be such that
we have that
$u^i(t_1,x)={\Bbb E}[e^{-r(\tau_x-t_1)}(K-S_{\tau_x}^{x,t_1,i})^+]$. Then
$\tau_x\geq t_2$ so $K e^{-r(t_2-t_1)}\geq K {\Bbb
  E}[e^{-r(\tau_x-t_1)}]\geq u^i(t_1,x)\geq (K-x)^+$. Letting $x\to 0^+$,
one deduces that $t_2=t_1$.
\end{adem}

Let us now prove some regularity properties of the pricing functions $u^i$.
\begin{alem}\label{operatorcontinuationregion}
Let $i\in\{1,\hdots,I\}$. Under the assumptions of Lemma \ref{defg}, 
 the function $u^i$ is continuous on the sets $[0,t_d^{i})\times {\Bbb R}_+$, $[t_d^{i},t_d^{i-1})\times {\Bbb R}_+$, $[t_d^{i-1},t_d^{i-2})\times {\Bbb R}_+$,...,$[t_d^{1},T]\times {\Bbb R}_+$  and for all $j\in\{1,\hdots,i\}$ and all $x$ outside
  the at most countable set of discontinuities of $D^j$, the limit $\lim_{t\to
  t_d^j-}u^i(t,x)$ exists and is equal to $u^i(t_d^j,x-D^j(x))$. Moreover, the exercise boundary $t\mapsto c^i(t)$ is upper-semicontinuous on $[0,T]$.\\
Last, for all points in the set $\{(t,x): t\in [0,T)\setminus\{t_d^k,1\leq k\leq i\},\, x>c^i(t)\}$ the partial derivatives $\partial_t u^i(t,x)$, $\partial_x u^i(t,x)$ and $\partial_{xx} u^i(t,x)$ exist and satisfy ${\cal A}u^i(t,\cdot)(x)+\partial_t u^i(t,x)=0$, and $u^i$ is $C^{1,2}$ on this set.
\end{alem}
\begin{adem}
Let us check the behaviour of $u^i$ as $t\to t_d^j-$ for $1\leq j\leq i\leq I$; the continuity of
$u^i$ follows from a similar but easier
argument.\\Since $S_{t_d^j}=S_{t^j_d-}-D^j(S_{t^j_d-})$, one has, using \eqref{lip} for the
inequality,
\begin{eqnarray*} |u^i(t,S_{t^j_d-})-u^i(t_d^j,S_{t^j_d-}-D^j(S_{t^j_d-}))|&=&|u^i(t,S_{t^j_d-})-u^i(t_d^j,S_{t_d^j})|
\\ &\leq&
|S_t-S_{t^j_d-}|+|u^i(t,S_{t})-u^i(t_d^j,S_{t_d^j})|.
\end{eqnarray*}
By continuity of the process
$(u^i(t,S_t))_{t\in[0,T]}$, which is ensured by Propositions \ref{snell} and \ref{pricfunc}, one deduces that a.s., $\lim_{t\to
  t_d^j-}u^i(t,S_{t_d^j-})=u^i(t_d^j,S_{t^j_d-}-D^j(S_{t^j_d-}))$. Since $S_{t_d^j-}$ admits a
positive density w.r.t. the Lebesgue measure on $(0,+\infty)$, $dx$ a.e. $\lim_{t\to
  t_d^j-}u^i(t,x)=u^i(t_d^j,x-D^j(x))$.  By
continuity of $x\mapsto u^i(t^j_d,x)$, the function $x\mapsto u^i(t^j_d,x-D^j(x))$ is
continuous outside the at most countable set of discontinuities of the
non-decreasing function $D^j$. With \eqref{lip}, one deduces that for
all $x$ outside this set, $\lim_{t\to
  t_d^j-}u^i(t,x)=u^i(t_d^j,x-D^j(x))$ and that $\forall x>c^i(t^j_d)$, $\liminf_{t\to t^j_d-}u^i(t,x)\geq u^i(t^j_d,x)>(K-x)^+$ which ensures that $\limsup_{t\to t^j_d-} c^i(t)\leq c^i(t^j_d)$. Since according to Lemma \ref{defg}, for $t\in [0,T]$, $\{x\geq 0:u^i(t,x)=(K-x)^+\}=[0,c^i(t)]$, the continuity properties of $u^i$ imply that $c^i$ is upper-semicontinuous on the sets $[0,t^i_d)$, $[t^i_d,t_d^{i-1})$, $[t^{i-1}_d,t_d^{i-2}),...,[t^1_d,T]$ and therefore on $[0,T]$.

Let $A_i=([0,T)\setminus\{t_d^k,1\leq k\leq i\}) \times {\Bbb
  R}_+$. By continuity of $u^i$ on $A_i$, the set $\{(t,x)\in A_i:\;x>c^i(t)\}$ is an open subset of $A_i$. 
Let $(t,x)\in A_i$ and $B$ be an open neighbourhood of $(t,x)$ with regular boundary $\partial B$ such that  $B$ is included in the connected component of $A_i$ which contains $(t,x)$. Define the stopping times
$\tau=\inf\{v\geq t: S_v^{x,t,i}\leq c^i(v)\}$ and $\tau_{B^c}=\inf\{v\geq
t: S_v^{x,t,i}\in B^c\}<\tau$. The flow property
for the Black-Scholes model without dividends implies that for $v\geq
\tau_{B^c}$,
$S_v^{x,t,i}=S_v^{ S_{\tau_{B^c}}^{x,t,i},\tau_{B^c},i}$ and
$\tau=\inf\{v\geq \tau_{B^c}: S_v^{ S_{\tau_{B^c}}^{x,t,i},\tau_{B^c},i}\leq
c^i(v)\}$. Using the strong Markov property for the third equality,
one deduces
\begin{align}
   u^i(t,x)=\E[ e^{-r(\tau-t)}(K-S_\tau^{x,t,i}
   )^+]&=\E[e^{-r(\tau_{B^c}-t)}\E[e^{-r(\tau-\tau_{B^c})}(K-S_\tau^{
     S_{\tau_{B^c}}^{x,t,i},\tau_{B^c},i })^+|{\cal F}_{\tau_{B^c}}]]\notag\\&=\E[e^{-r(\tau_{B^c}-t)}u^i(\tau_{B^c},S_{\tau_{B^c}}^{x,t,i})].\label{conditmarkov}
\end{align}
Let $f(s,x)$ be a solution to the Dirichlet problem where $\partial_s f +{\cal A}f=0$ on $B$ and $f=u^i$ on $\partial B$. By Theorem 3.6.3. in \cite{Friedman} this function $f$ is $C^{1,2}$ in $B$ and continuous on $\bar{B}$. But then \begin{align*}
   u^i(t,x)&={\Bbb E}[e^{-r(\tau_{B^c}-t)} u^i(\tau_{B^c},S_{\tau_{B^c}}^{x,t,i})]=
 {\Bbb E}[e^{-r(\tau_{B^c}-t)} f(\tau_{B^c},S_{\tau_{B^c}}^{x,t,i})]\\&=f(t,x)+{\Bbb E}\int_t^{\tau_{B^c}} e^{-r(s-t)}(\partial_s f +{\cal A}f)(s,S^{x,t,i}_s)ds=f(t,x)
\end{align*} by optional sampling so $u^i=f$ on $B$ and therefore its partial derivatives exist in $(t,x)$ and they satisfy $\partial_t u^i(t,x)+{\cal A}u^i(t,\cdot)(x)=0$.
\end{adem}

The characterization of the restriction of $u^i$ to $[0,t_d^{i})\times\R_+$ as
the pricing function of an American option in the Black-Scholes model
without dividends, as stated in the next proposition, is the key to the study
of the exercise boundaries $c^i(t)$ performed in the following sections.
\begin{aprop}\label{reducmodel}
Under the assumptions of Lemma \ref{defg}, we have for all $0\leq i\leq I$,
\[
 \forall (t,x)\in [0,t_d^{i}) \times\R_+,\;u^i(t,x) = \sup_{\tau \in {\cal T}_{[0,t_d^i-t]}} {\Bbb E}[ e^{-r\tau}(\, (K-\S_{\tau}^x)^+1_{\{\tau<t_d^i-t\}}+g^i(\S_{t_d^i}^x)1_{\{\tau=t_d^i-t\}}  \, ) ].
\] where $g^0(x)\stackrel{\rm def}{=}(K-x)^+$ and $g^i(x)\stackrel{\rm def}{=}u^{i-1}(t_d^i,x-D^i(x))$ for $i\geq 1$, and the supremum is attained for 
$\tau=\inf\{s\in [0,t_d^i-t): \S^x_s\leq c^i(t+s)\}\wedge t_d^i-t$ (with the convention that
$\inf\emptyset=+\infty$).
Moreover, for all $0\leq j\leq i$  and $t\in [t_d^{j+1},T]$, we have that $c^i(t)=c^j(t)$ and $u^i(t,x)=u^j(t,x)$ for all positive $x$.
 \end{aprop}
\begin{adem}
For $i=0$ the statement is trivial so assume $i\geq 1$.
The last statement of the proposition is obvious because when $0\leq j\leq i$, the optimal stopping problems in proposition \ref{pricfunc} which define the values $u^i(t,x)$ and $u^j(t,x)$ and the values $c^i(t)$ and $c^j(t)$ are the same
for $t\geq t_d^{j+1}$ and $x\geq 0$ because we then have that $S_v^{t,x,i}=S_v^{t,x,j}$ for $v\in[t,T]$.
Take $t\in [0,t_d^{i})$ and $x\geq 0$ and define $\tau_x=\inf\{v\geq t:
S_v^{x,t,i}\leq c^i(v)\}$. 
Arguing like in the derivation of \eqref{conditmarkov}, one easily checks that
\begin{eqnarray*}
\lefteqn{
\E\left[e^{-r(\tau_x-t)}(K-S^{x,t,i}_{\tau_x})^+1_{\{\tau_x\geq
  t_d^i\}}\right]=   \E\left[e^{-r(t_d^i-t)}u^i(t_d^i,S^{x,t,i}_{t_d^i})1_{\{\tau_x\geq
  t_d^i\}}\right]}
  &&\\&=& \E\left[e^{-r(t_d^i-t)}u^{i-1}(t_d^i,S^{x,t,i}_{t_d^i})1_{\{\tau_x\geq
  t_d^i\}}\right]
  =\E\left[e^{-r(t_d^i-t)}g^i(S^{x,t,i}_{t^i_d-})1_{\{\tau_x\geq t_d^i\}}\right]
  \end{eqnarray*} where we used the previous result for $j=i-1$ to obtain the second equality. We thus
deduce that 
\begin{align*} u^i(t,x)&=\E\left[e^{-r(\tau_x-t)}(K-S^{x,t,i}_{\tau_x})^+1_{\{\tau_x<t_d^i\}}+e^{-r(t_d^i-t)}g^i(S^{x,t,i}_{t^i_d-})1_{\{\tau_x\geq t_d^i\}}\right]\\
&=\E\left[e^{-r\tau}(K-\S^{x}_{\tau})^+1_{\{\tau<t_d^i-t\}}+e^{-r(t_d^i-t)}g^i(\S^{x}_{t_d^i-t})1_{\{\tau=t_d^i-t\}}\right],
\end{align*}
when $\tau=\inf\{s\in [0,t_d^i-t): \S^x_s\leq c^i(t+s)\}\wedge t_d^i-t$.\\
Let now $\tau$ be any stopping time in ${\cal T}_{[0,t_d^i-t]}$. For $f:C([0,t_d^i-t],\R)\to
[0,t_d^i]$ such that $\tau=f(W_s,0\leq s\leq t_d^i-t)$, the random variable
\begin{equation*}
   \tau_x\stackrel{\rm def}{=}\begin{cases}
      t+f(W_s-W_t,t\leq
s\leq t_d^i)\mbox{ if }t+f(W_s-W_t,t\leq
s\leq t_d^i)<t_d^i\\
\inf\{s\geq t_d^i:S^{x,t,i}_s\leq c^i(s)\}\mbox{ otherwise}
   \end{cases}
\end{equation*}
belongs to ${\cal T}_{[t,T]}$ and  is such that
\begin{align*}
   \E&\left[e^{-r\tau}(K-\S^{x}_{\tau})^+1_{\{\tau<t_d^i-t\}}+e^{-r(t_d^i-t)}g^i(\S^{x}_{t_d^i-t})1_{\{\tau=t_d^i-t\}}\right]\\
&=\E\left[e^{-r(\tau_x-t)}(K-S^{x,t,i}_{\tau_x})^+1_{\{\tau_x<t_d^i\}}+e^{-r(t_d^i-t)}u^{i-1}(t_d^i,S^{x,t,i}_{t_d^i})1_{\{\tau_x\geq
    t_d^i\}}\right]\\
&=\E\left[e^{-r(\tau_x-t)}(K-S^{x,t,i}_{\tau_x})^+\right]\leq u^i(t,x).
\end{align*}
\end{adem}

 This result shows that it is natural to consider the case with only one dividend date first and then use the results  to
generalize to multiple dividend dates. This will allow us to prove the following result for the multiple dividend problem:

\begin{thm}\label{multiple}
Let for all $1\leq j\leq I$ the dividend functions $D^j$ be non-negative, non-decreasing
and such that $x\in\R_+\mapsto x-D^j(x)$ is non-negative and
non-decreasing. Then for all $1\leq i\leq I$ the exercise boundaries $c^i$ are strictly positive and locally bounded away from zero
on $[t_d^{i+1},t_d^i)$. If $D^i$ is positive, then $\lim_{t\to {t^i_d}-}c^i(t)=0$ with $c^i(t)\leq rK(t^i_d-t)\inf_{x>0}\frac{x}{D^i(x)}+o(t^i_d-t)$ as $t\to{t^i_d}-$ when $D^i$ is also concave. Moreover, if for all $1\leq j\leq i$ we have $D^j(x)=(1-\rho_j)x$ for some $\rho_j\in (0,1)$ then
\begin{itemize}
\item for all $t\in [0,T]$ the value function $u^i(x,t)$ is convex in $x$, 
\item $c^i$ is right-continuous on $[0,T]$ and $\forall t\in [0,T)$, $\partial_x u^i(t,c^i(t)^+)=-1$ i.e. the smooth contact property holds, and
\item there exist $\varepsilon^i>0$ such that on $ (t_d^i-\varepsilon^i,t_d^i)$, the function $c^i$ is continuous and non-increasing with  
$c^i(t)\sim rK(t_d^i-t)/(1-\rho_i)$ as $t\to t_d^i-$.
\end{itemize}
\end{thm}

 The proof for this Theorem can be found in the Appendix. It is based on the stronger results that we will  prove for the single dividend case in this section and the next two sections. Remember that in the single dividend case we use the shorthand notation $u(t,x)=u^1(t,x)$, $g(x)=g^1(x)$, $D(x)=D^1(x)$ and $t_d=t_d^1$  and that $\U(t,x)=u^0(t,x)$ and $\bar{c}(t)=c^0(t)$ are used for the case when no dividends are present. We will also write $S^{x,t}$ for $S^{x,t,1}$ now that $I=1$.
 
 We first derive some properties of the function $g(x)=\U(t_d,x-D(x))$. 

\begin{alem}\label{propg}
Assume that $D$ is a non-negative concave function such that
$x-D(x)$ is non-negative.  Then $D$ is continuous, non-decreasing and
such that $x-D(x)$ is non-decreasing. Let $D'_-(x)$ and $D''(dx)$
respectively denote the left-hand
derivative of $D$ and the non-positive Radon measure equal to the second order
distribution derivative of $D$ on $(0,+\infty)$. The function $g$ is continuous,
non-increasing and $g(x)\geq (K-x)^+$ for all $x\geq 0$.
The function 
$$\gamma(x)\stackrel{\rm def}{=}\frac{\sigma^2
  x^2}{2}(1-D_-'(x))^2\partial_{22}\U(t_d,x-D(x))+rx(1-D_-'(x))\partial_{2}\U(t_d,x-D(x))-r\U(t_d,x-D(x))$$
where, by convention, $\partial_{22}\U(t_d,\bar{c}(t_d))=0$,
 is not greater than $-rK$ on $(0,x^\star)$ where $x^\star\stackrel{\rm
   def}{=}\sup\{x:x-D(x)<\bar{c}(t_d)\}>0$, and globally bounded.
   
   If $g$ is convex, then there is a constant $\rho\in[0,1]$
such that $g(x)=K-\rho x$ and $D(x)=(1-\rho)x$ for $x<x^\star$, the second order distribution
 derivative of $g$ admits a density $g''$ w.r.t. the Lebesgue measure
 and ${\cal A}g(x)$ is equal to $-rK$
 on $(0,x^\star)$ and $dx$ a.e. on $(x^\star,+\infty)$, ${\cal A}g(x)\geq-rK$.
\end{alem}
To prove this lemma, we need the following properties of the pricing
function $\U$ in the model without dividends.
\begin{alem}\label{dertemps} For the case without dividends we have that
  the partial derivatives $\partial_t \U(t,x)$, $\partial_x \U(t,x)$ and
  $\partial_{xx} \U(t,x)$ exist and $\partial_t
  \U(t,x)+{\cal A}\U(t,\cdot)(x)=0$ for all $t\in [0,T)$ and
  $x>\bar{c}(t)$. Moreover, $\forall t\in [0,T]$, $x\mapsto \U(t,x)$ is
  convex and $C^1$ on $\R_+$. Last,
$$\forall t\in [0,T),\;\forall x>\bar{c}(t),\;\partial_t \U(t,x)\geq -\frac{e^{-r(T-t)}\sigma K}{2\sqrt{2\pi
       (T-t)}}\exp\left(-\frac{(\log(K/x)-(r-\frac{\sigma^2}{2})(T-t))^2}{2\sigma^2(T-t)}\right).$$
\end{alem}
Before proving these Lemmas, let us give some examples of functions $g$
obtained for different choices of the dividend function $D$.

\begin{figure}[ht]
 \begin{center}
\psfrag{x}{\Large $x$}
\psfrag{K}{\Large $K$}
\psfrag{D}{\Large $D$}
\psfrag{ctd}{\Large $\bar{c}(t_d)$}
\psfrag{x0}{\Large $x_0$}
\psfrag{u(td,x)}{\Large $\bar{u}(t_d,x)$}
\psfrag{const div D=1}{\Large Const div $D=1$}
\psfrag{prop div rho=0.55}{\Large Prop div $\rho=0.75$}
\psfrag{convex example}{\Large Convex example}\scalebox{0.75}{\includegraphics{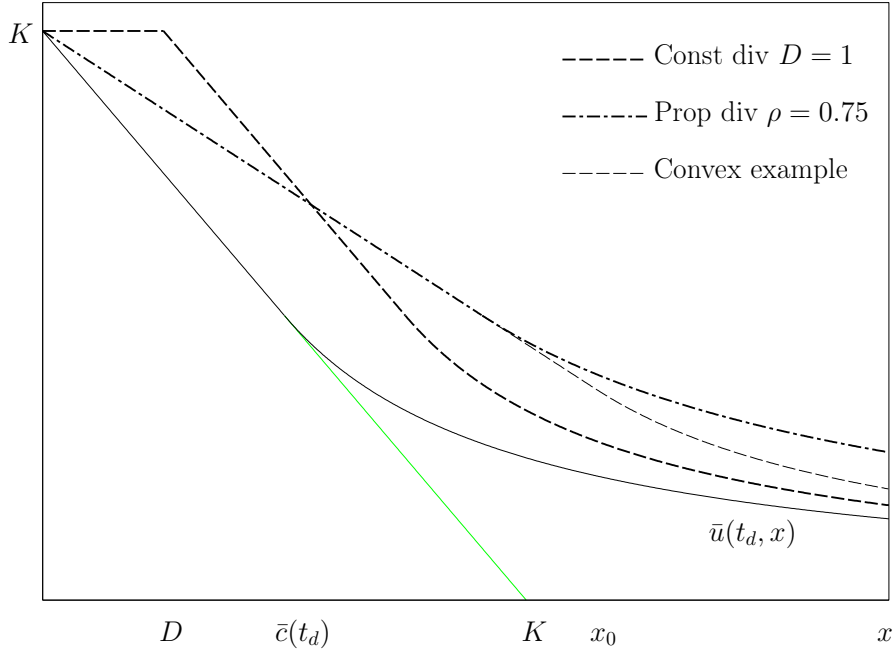}}
 \end{center}
\caption{\label{fig:exg}Examples of functions $g$}
 \end{figure}

{\bf Examples of functions $g$ :}\begin{itemize}
   \item In the {\sl constant} dividend case, $x^\star=\bar{c}(t_d)+D$ and the function $g$
is equal to $K$ on $[0,D]$ and to $K+D-x$ for
$x\in(D,x^\star)$, $C^1$ on $[0,D)\cup(D,+\infty)$ with $g'$ taking its
values in $[-1,0]$, $C^2$ on
$[0,D)\cup(D,x^\star)\cup(x^\star,+\infty)$ and such that 
${\cal A}g(dx)=\gamma(x)dx-\frac{\sigma^2 D^2}{2}\delta_D(dx)$ where
$\gamma$ is equal to $-rK$ on $(0,D)$ and to $-r(K+D)$ on $(D,x^\star)$.
\item In the {\sl proportional} dividend case, $x^\star=\bar{c}(t_d)/\rho$ and
  $g(x)=\U(t_d,\rho x)$ is convex, $C^1$ with $g'$ taking its values in
$[-\rho,0]$ and $C^2$ on $[0,x^\star)\cup(x^\star,+\infty)$.
\item The {\sl proportional} dividend case provides an example of a non-negative
   concave function $D$ such that
$x-D(x)$ is non-negative which leads to a convex function $g$. This
example is not unique. For instance, let $\rho\in
(0,1)$. The
function $y\mapsto \U(t_d,y)$ is convex positive nonincreasing and such
that $\lim_{y\to +\infty}\U(t_d,y)=0$. So it is continuous and
decreasing and admits an inverse $V(t_d,.):(0,K]\to [0,+\infty)$. For
$x\in (\bar{c}(t_d)/\rho,K/\rho)$, we set $d(x)=x-V(t_d,K-\rho x)$. The
continous function
$d'(x)=1+\rho/\partial_2\U(t_d,V(t_d,K-\rho x))$ is non-increasing on
$(\bar{c}(t_d)/\rho,K/\rho)$ by the non-increasing property of both $V(t_d,.)$ and
$-\partial_2\U(t_d,.)$ and the positivity of this last function. It tends
respectively to $1-\rho$ and $-\infty$ as $x\to \bar{c}(t_d)/\rho$ and $x\to K/\rho$. Let
$x_0=\sup\{x\in(\bar{c}(t_d)/\rho,K/\rho):d'(x)\geq 0\}$. One has $d'(x_0)=0$
which also writes $\partial_2\U(t_d,x_0-d(x_0))=-\rho$. The function
\begin{equation*}
   D(x)=\begin{cases}(1-\rho)x\mbox{ for }x\in [0,\bar{c}(t_d)/\rho]\\
     d(x\wedge x_0)\mbox{ for }x>\bar{c}(t_d)/\rho
   \end{cases}
\end{equation*}
is non-negative, concave and such that $x-D(x)$ is non-negative. The
convexity of $x\mapsto \U(t_d,x)$ combined with the equality
$\partial_2\U(t_d,x_0-d(x_0))=-\rho$ implies that
\begin{equation*}
   g(x)=\begin{cases}K-\rho x\mbox{ for }x\in [0,x_0]\\
     \U(t_d,x-d(x_0))\mbox{ for }x>x_0
   \end{cases}
\end{equation*}
is convex.
\end{itemize}
Figure \ref{fig:exg} illustrates the construction of the function
  $g$ from $x\mapsto \U(t_d,x)$ on the three previous examples of
  dividend functions.\begin{adem}[of Lemma \ref{propg}]
Since the concave function $D$ is non-negative, it is continuous and
non-decreasing. And since $x-D(x)$ is non-negative, $D(0)=0$. The convex
function $x-D(x)$ being non-negative and equal to $0$ for $x=0$, is non-decreasing.
Since $x\mapsto \U(t_d,x)$ is continuous, non-increasing and not
smaller than $(K-x)^+$, the same properties hold for $g$. \\
For $x\in(0,x^\star)$,
$\gamma(x)=rx(D'_-(x)-1)-r(K-x+D(x))=-rK-r(D(x)-xD'_-(x))$. 
By concavity of $D$, 
\begin{align}\label{minod-xd'}
   \forall x>0,\;D(x)-xD'_-(x)\geq D(0)=0.
\end{align}So $\gamma$ is not greater than $-rK$ on $(0,x^\star)$. The
constant $x^\star$ is infinite if and only if $D$ is the identity
function and then $\gamma$ is constant and equal to $-rK$. When
$x^\star<+\infty$, $\gamma$ is bounded from below by $-r(K+D(x^\star))$
on $(0,x^\star)$. Moreover, since $D$ is concave, continuous and $D(0)=0$, 
\begin{align}
   \forall x>x^\star,\;\frac{D(x)}{x}\leq
\frac{D(x^\star)}{x^\star}=\frac{x^\star-\bar{c}(t_d)}{x^\star}\mbox{ and }x-D(x)\geq \frac{x \bar{c}(t_d)}{x^\star}>\bar{c}(t_d)\label{contd}.
\end{align} One has
\begin{align}\label{eq:phi2}
   \gamma(x)-{\cal A}\U(t_d,.)(x-D(x))&=\frac{\sigma^2}{2}\partial_{22}\U(t_d,x-D(x))[x^2(1-D'_-(x))^2-(x-D(x))^2]\\
   &+r(D(x)-xD'_-(x))\partial_2\U(t_d,x-D(x))\nonumber
\end{align}
where the last term is non-positive by \eqref{minod-xd'} and since
$\partial_2\U\leq 0$.
Define 
$M=\sup_{x>\bar{c}(t_d)}{\cal A}\U(t_d,.)(x)$ which is finite by Lemma \ref{dertemps}.
Since
$\U(t_d,x)-x\partial_x \U(t_d,x)$ is non-increasing by convexity of
$x\mapsto \U(t_d,x)$ and equal to $K$ on $[0,\bar{c}(t_d))$, one deduces 
\begin{align}\label{majodersecU}
   \forall x>\bar{c}(t_d),\;\partial_{xx}\U(t_d,x)\leq
\frac{2(M+rK)}{\sigma^2 x^2}.
\end{align}
With $x-D(x)$, which is larger than $\bar{c}(t_d)$, substituted in
\eqref{majodersecU}, and using \eqref{contd} and $D'_{-}(x)\in[0,1]$, one
concludes that when $x^\star<+\infty$,
\begin{align*}
\forall x>x^\star,\;\gamma(x)&\leq
M+(M+rK)\frac{{x^\star}^2-\bar{c}(t_d)^2}{\bar{c}(t_d)^2}.
\end{align*}
For $x>x^\star$, since $xD'_-(x)\partial_2\U(t_d,x-D(x))$ and
$\partial_{22}\U(t_d,x-D(x))[x^2(1-D'_-(x))^2-(x-D(x))^2]$
are non-negative and ${\cal A}\U(t_d,.)(x-D(x))=-\partial_t
\U(t_d,x-D(x))>0$,  we have by (\ref{eq:phi2}),
\begin{align*}
  \gamma(x)&\geq  rD(x)\partial_2\U(t_d,x-D(x))\geq
  r\frac{x^\star-\bar{c}(t_d)}{\bar{c}(t_d)}(x-D(x))\partial_2\U(t_d,x-D(x))\\&=r\frac{x^\star-\bar{c}(t_d)}{\bar{c}(t_d)}\left(-K+\int_{\bar{c}(t_d)}^{x-D(x)}y\partial_{22}\U(t_d,y)dy+\U(t_d,x-D(x))\right)
\geq -rK\frac{x^\star-\bar{c}(t_d)}{\bar{c}(t_d)},
\end{align*}
where we used that $D(x)\leq (x-D(x))(x^*-\bar{c}(t_d))/\bar{c}(t_d)$ by
\eqref{contd} for the second inequality and the smooth fit property
$\partial_2\U(t_d,\bar{c}(t_d))=-1$ and a partial integration
for the equality.\par Last, assume that $g$ is convex. If $g'_+$ and $D'_+$ respectively denote the right-hand derivatives of $g$ and $D$, one has $g'_+(x)-g'_-(x)=-\partial_2\U(t_d,x-D(x))(D'_+(x)-D'_-(x))$ and 
since
$\partial_2 \U$ is negative and $D'_+-D'_-$ non-positive, the
right-hand-side of this equality is non-positive and the left-hand-side
is non-negative. So both are zero and
the functions $g$ and $D$ are $C^1$ with
$g'(x)=\partial_2\U(t_d,x-D(x))(1-D'(x))$. The first factor in the
right-hand-side being
globally continuous and $C^1$ on $(0,x^\star)\cup(x^\star,+\infty)$, one deduces that the
distribution derivative of $g'$ is equal to
$\partial_{22}\U(t_d,x-D(x))(1-D'(x))^2dx-\partial_2\U(t_d,x-D(x))D''(dx)$.
This measure being non-negative by convexity of $g$, $D''$ is absolutely
continuous with respect to the Lebesgue measure and so is the second
order distribution derivative of $g$. For $x<x^\star$,
$g'(x)=D'(x)-1$ where the left-hand-side is non-decreasing and the
right-hand-side non-increasing. So there is a constant $\rho\in[0,1]$
such that $g(x)=K-\rho x$ and $D(x)=(1-\rho)x$ for $x<x^\star$. As a consequence
$x^\star=\bar{c}(t_d)/\rho$ and ${\cal A}g(x)=rxg'(x)-rg(x)=-rK$ on
$(0,x^\star)$. The convexity of $g$ implies that $rxg'(x)-rg(x)$ is
non-decreasing and therefore that $dx$ a.e. on $(x^\star,+\infty)$,
${\cal A}g(x)=\frac{\sigma^2x^2}{2}g''(x)+rxg'(x)-rg(x)\geq -rK$.
\end{adem}

\begin{adem}[of Lemma \ref{dertemps}] The proof of the first statement
  is similar to the one of the last statement in Lemma
  \ref{operatorcontinuationregion}. Moreover, $x\mapsto
  \U(t,x)=\sup_{\tau\in{\cal T}_{[0,T-t]}}\E\left(e^{-r\tau}(K-xe^{\sigma
    W_\tau+(r-\frac{\sigma^2}{2})\tau})^+\right)$ is convex as the supremum of
  convex functions. We refer for instance to Lemma 7.8 in Section 2.6
  \cite{KaratzasShreve} for the continuous differentiability property of
  this function.\\
Let $0\leq s\leq t\leq T$, $x>0$, and take $\tau\in{\cal T}_{[0,T-s]}$ such that $\U(s,x)=\E(e^{-r\tau}(K-\S^x_\tau)^+)$ and
   $\tilde{\tau}=\tau\wedge(T-t)$. One has
\begin{align*}
   \U(t,x)\geq \E\left(e^{-r\tilde{\tau}}(K-\S^x_{\tilde{\tau}})^+\right)=\U(s,x)-\E\left(1_{\{\tau>T-t\}}\left(e^{-r\tau}(K-\S^x_\tau)^+-e^{-r(T-t)}(K-\S^x_{T-t})^+\right)\right)
\end{align*}
By Tanaka's formula, when $\tau>T-t$, 
$$(K-\S^x_\tau)^+=(K-\S^x_{T-t})^+-\int_{T-t}^\tau 1_{\{\S^x_v\leq
  K\}}(\sigma \S^x_v dW_v+r
\S^x_vdv)+\frac{1}{2}(L^K_\tau(\S^x)-L^K_{T-t}(\S^x)).$$
One deduces that
$$\U(t,x)\geq
\U(s,x)-\frac{e^{-r(T-t)}}{2}\E(L^K_{T-s}(\S^x)-L^K_{T-t}(\S^x))=\U(s,x)-\frac{e^{-r(T-t)}\sigma^2
    K^2}{2}\int_s^t p(T-v,K)dv.$$
\end{adem}

\section{Limit behaviour and monotonicity of the exercise boundary as
  $t\to t_d^-$}

Using the results in the previous section, we first
check that $c(t)$ tends to $0$ as $t\to t_d^-$ if $D$ is positive
(i.e. $\forall x>0$, $D(x)>0$).

\begin{alem}\label{limnul}Let $D$ be a non-negative and non-decreasing
  function s.t. $x\mapsto x-D(x)$ is non-negative and
  non-decreasing. 

Assume moreover that there exists a $d_0\geq 0$ such that $D$ is zero on $[0,d_0]$ and positive on $]d_0,\infty[$, then we have $\limsup_{t\to t_d^-}c(t)\leq d_0\wedge \bar{c}(t_d)$. When $d_0=0$ i.e. $D$ is positive, then $\lim_{t\to t_d-}c(t)=0$ and
\begin{itemize}
\item if $D$ is such that $\frac{x}{D(x)}$ admits a finite limit as $x\to 0^+$ then $c(t)\leq rK(t_d-t)\lim_{x\to 0^+}\frac{x}{D(x)}+o(t_d-t)$ as $t\to t_d^-$,
\item if $D$ is concave, $g$ is convex and the constant $\rho$ such that, according to Lemma
  \ref{propg}, $\forall x\in (0,x^\star)$, $D(x)=(1-\rho)x$ belongs to $(0,1)$ then
  $\forall t\in [0,t_d),\;c(t)<\frac{1-e^{-r(t_d-t)}}{1-\rho}K$. When $\rho=0$ i.e. $D$ is the identity function, then $\forall t\in [0,t_d),\;c(t)\leq (1-e^{-r(t_d-t)})K$.
\end{itemize}
\end{alem}
Note that when $D$ is postive and concave, then $\frac{x}{D(x)}$ admits a finite limit as $x\to 0^+$ which is equal to $\inf_{x>0}\frac{x}{D(x)}$.
\begin{adem}
  Suppose that  $\limsup_{t\to t_d^-}c(t) > d_0\wedge \bar{c}(t_d)$, then there exists a $y> 0$ and a sequence $(t_n)_{n\in {\Bbb N}}$ such that $t_n\uparrow t_d$ with
  $c(t_n)>y>d_0\wedge \bar{c}(t_d)$ and since $c(t_n)\leq K$ we have $y<K$ and we may choose $y$ such that it is not one of the countably many discontinuity points of $D$. 
  Then $K-y=u(t_n,y)$ for all $t_n$ and taking the limit and applying Lemma \ref{operatorcontinuationregion} gives that $K-y=\bar{u}(t_d,y-D(y))$ but either $y>d_0$ and then $\bar{u}(t_d,y-D(y))\geq (K-y+D(y))^+=K-y+D(y)>K-y$ or $y>\bar{c}(t_d)$ and then $\bar{u}(t_d,y-D(y))\geq \bar{u}(t_d,y)>K-y$ so in both cases we get a contradiction. 
Assume that $D$ is such that $\mu\stackrel{\rm def}{=}\lim_{x\to 0^+}\frac{x}{D(x)}$ exists and is finite. Since both $D(x)$ and $x-D(x)$ are non-negative, necessarily $\mu\geq 1$. 
 For $(t,x)\in[0,t_d)\times\R_+^*$,
\begin{align*}
   u(t,x)&\geq \E(e^{-r{(t_d-t)}}g(\S^x_{t_d-t}))\geq
   \E\left(e^{-r{(t_d-t)}}\left(K-\S^x_{t_d-t}+\frac{D(\S^x_{t_d-t})}{\S^x_{t_d-t}}\S^x_{t_d-t}\right)\right)\\
&\geq e^{-r{(t_d-t)}}K-x+\inf_{0<y\leq 4rK\mu(t_d-t)}\frac{D(y)}{y}\left(x-\E\left(e^{-r{(t_d-t)}}\S^x_{t_d-t}1_{\{\S^x_{t_d-t}>4rK\mu(t_d-t)\}}\right)\right)\\
&\geq e^{-r{(t_d-t)}}K+\left(\inf_{0<y\leq 4rK\mu(t_d-t)y}\frac{D(y)}{y}-1\right)x-x N\left(\frac{\log(\frac{x}{4rK\mu (t_d-t)})+(r+\frac{\sigma^2}{2})(t_d-t)}{\sigma\sqrt{t_d-t}}\right).
\end{align*}
For $x\leq 2rK\mu(t_d-t)$ and $(t_d-t)\leq \frac{\log(2)}{2r+\sigma^2}$,  $\frac{\log(\frac{x}{4rK\mu (t_d-t)})+(r+\frac{\sigma^2}{2})(t_d-t)}{\sigma\sqrt{t_d-t}}\leq -\frac{\log(2)}{2\sigma\sqrt{t_d-t}}$ which implies
$N\left(\frac{\log(\frac{x}{4rK\mu (t_d-t)})+(r+\frac{\sigma^2}{2})(t_d-t)}{\sigma\sqrt{t_d-t}}\right)\leq \frac{2\sigma\sqrt{t_d-t}}{\sqrt{2\pi}\log(2)}e^{-\frac{\log^2(2)}{8\sigma^2(t_d-t)}}$.
With $\lim_{t\to t_d^-}\inf_{0<y\leq 4rK\mu(t_d-t)}\frac{D(y)}{y}=\frac{1}{\mu}$, one deduces that, as $t\to t_d^-$, for $x\leq 2rK\mu(t_d-t)$, $u(t,x)\geq K-x+\left(\frac{x}{\mu}-rK(t_d-t)\right)+o(t_d-t)$ where the $o(t_d-t)$ does not depend on $x$. One easily deduces the desired upper-bound for $c(t)$.\\


When $g$ is also convex, according to Lemma \ref{propg}, either $D$ is the identity function and $g$
is constant and equal to $K$ or there is a constant $\rho \in (0,1)$
such that $D(x)=(1-\rho)x$ for $x\in (0,\bar{c}(t_d)/\rho]$. In the latter
case, one has
$g(x)=K-\rho x$ for $x\in (0,\bar{c}(t_d)/\rho]$ and $g(x)\geq (K-\rho x)^+$ for $x>\bar{c}(t_d)/\rho$.
As a consequence,
$\E(e^{-r{(t_d-t)}}g(\S^x_{t_d-t}))>\E(e^{-r(t_d-t)}(K-\rho
\S^x_{t_d-t}))=e^{-r(t_d-t)}K-\rho x$.
One deduces that when $x\geq\frac{1-e^{-r(t_d-t)}}{1-\rho}K$,
$u(t,x)>K-x$ which implies that
$c(t)<\frac{1-e^{-r(t_d-t)}}{1-\rho}K$. When $D$ is the identity function, the inequality is obvious.
\end{adem}

We now obtain monotonicity of the exercise boundary in a left-hand
neighbourhood of the dividend date $t_d$.
\begin{aprop}\label{propcle}If $D$ is a positive concave function such that $x-D(x)$ is non-negative, there exists a constant $\varepsilon >0$ such that for $x\in
   (0,\varepsilon)$, $t\mapsto u(t,x)$ is non-decreasing on
 $(t_d-\varepsilon,t_d)$. 
 Moreover, we have  for all $t\in [0,t_d)$ and all $x>c(t)$ that
 \begin{eqnarray}\label{minodertemps}
\partial_t u(t,x)&\geq&
   -e^{-r(t_d-t)}\sup_{y>0}\gamma^+(y)\\
   \frac{\sigma^2
     x^2}{2}\partial_{xx}u(t,x)&\leq& e^{-r(t_d-t)}\sup_{y>0}\gamma^+(y)+r(x+K).
\label{majodersec} \end{eqnarray}
Last, for any $t\in [0,t_d)$ such that $c(t)>0$, $\forall
x>c(t),\int_{c(t)}^x|\partial_{xx}u(t,y)|dy<+\infty$ and $x\mapsto \partial_x
u(t,x)$ admits a right-hand limit $\partial_x
u(t,c(t)^+)\in[-1,0]$ as $x\to c(t)^+$.
\end{aprop}
One easily deduces the following Corollary.
\begin{acor}\label{frontdeccag}
If the dividend function $D$ is non-negative, non-decreasing
and such that $x\in\R_+\mapsto x-D(x)$ is non-negative and
non-decreasing, then the exercise boundary does not vanish on $[0,T]$. Moreover, for all $t\in[0,t_d)$, $\inf_{s\in[0,t]}c(s)>0$.\\
   If $D$ is a positive concave function such that $x-D(x)$ is
   non-negative, then $t\mapsto c(t)$ is non-increasing and left-continuous on $(t_d-\varepsilon,t_d)$. Moreover, $c(t)\sim rK(t_d-t)\inf_{x>0}\frac{x}{D(x)}$ as $t\to t_d^-$.
\end{acor}\begin{arem}
In contrast to the result of Corollary \ref{frontdeccag}, we notice that in the alternative model formulation known as the Escrowed model $$S_t=(S_0-De^{-rt_d})e^{\sigma W_t+(r-\frac{\sigma^2}{2})t}+De^{-r(t_d-t)}1_{\{t<t_d\}}$$ where $D$ is a positive constant, the boundary is actually equal to $0$ on a left-hand neighbourhood of $t_d$. Indeed, reasoning like in the proof of Proposition \ref{reducmodel}, one can check that for $(t,x)\in (0,t_d)\times\R_+$, the value function in this model is 
$$u(t,x)=\sup_{\tau\in{\cal T}_{[0,t_d-t]}}\E\left[e^{-r\tau}((K-De^{-r(t_d-\tau)}-\bar{S}^y_\tau)^+1_{\{\tau<t_d-t\}}+\bar{u}(t_d,\bar{S}^y_{t_d-t})1_{\{\tau=t_d-t\}})\right]$$
where $y=x-De^{-r(t_d-t)}$. Since $$\E\left(e^{-r(t_d-t)}\bar{u}(t_d,\bar{S}^y_{t_d-t})\right)\geq \E\left(e^{-r(t_d-t)}(K-\bar{S}^y_{t_d-t})^+\right)\geq (Ke^{-r(t_d-t)}-y)^+,$$
early exercise is never optimal when $K-De^{-r(t_d-t)}< Ke^{-r(t_d-t)}$ i.e. $t_d-t<\frac{1}{r}\log\left(\frac{K+D}{K}\right)$.
\label{Escrowed}\end{arem}
\begin{adem}[of Corollary \ref{frontdeccag}]
For $t\in[t_d,T]$, $c(t)$ is larger than the exercise boundary $\frac{2rK}{\sigma^2+2r}$ of the perpetual Put in the Black-Scholes model without dividends. For $(t,x)\in[0,t_d)\times\R_+$, by Proposition \ref{reducmodel}, the pricing function $u(t,x)$ is smaller than the one corresponding to the identity dividend function. Therefore for $t\in[0,c(t_d))$, $c(t)$ is larger than the associated boundary. For the identity dividend function the function $\gamma$ is constant and equal to $-rK$ so that the exercise boundary is non-increasing on $[0,t_d)$ by \eqref{minodertemps} and therefore does not vanish by Lemma \ref{defg}.

Let us now assume that $D$ is a positive concave function such that $x-D(x)$ is
   non-negative. The monotonicity of $c$ is a consequence of Proposition \ref{propcle} and the left continuity
then follows from the upper-semicontinuity.
Let us now assume that $c(t)$ is not equivalent to $rK\mu (t_d-t)$ where $\mu=\inf_{x>0}\frac{x}{D(x)}$ as $t\to t_d^-$ and obtain a contradiction. Because of the upper-bound stated in Lemma \ref{limnul}, this implies the existence of a constant $\tilde{\mu}\in (0,\mu)$ and a sequence $(t_n)_{n\in\N}$ in $(t_d-\varepsilon,t_d)$ such that $\lim_{n\to\infty}t_n=t_d$ and $\forall n\in\N$, $c(t_n)\leq rK\tilde{\mu}(t_d-t_n)$. For $n\in\N$, let $x_n=\frac{\mu+\tilde{\mu}}{2}rK(t_d-t_n)$ and $\tau_n=\inf\{s\in [0,t_d-t_n):\S^{x_n}_s\leq c(t_n+s)\}\wedge(t_d-t_n)$ denote the optimal stopping time starting from $x_n$ at time $t_n$. One has
\begin{align}
   u(t_n,x_n)&\leq K\P\left(\exists s\in [0,t_d-t_n):\S^{x_n}_s\leq c(t_n)\right)+\E\left(e^{-r(t_d-t_n)}g(\S^{x_n}_{t_d-t_n})\right)\notag\\
&\leq K\P\left(\sigma\inf_{s\in[0,t_d-t_n]}W_s\leq \log(\frac{2\tilde{\mu}}{\mu+\tilde{\mu}})+(r-\frac{\sigma^2}{2})^-(t_d-t_n)\right)+\E(e^{-r(t_d-t_n)}g(\S^{x_n}_{t_d-t_n}))\notag\\
&= K\P\left(|W_{t_d-t_n}|\geq \frac{1}{\sigma}\left(\log(\frac{\mu+\tilde{\mu}}{2\tilde{\mu}})-(r-\frac{\sigma^2}{2})^-(t_d-t_n)\right)\right)+\E(e^{-r(t_d-t_n)}g(\S^{x_n}_{t_d-t_n}))\notag\\
&=\E(e^{-r(t_d-t_n)}g(\S^{x_n}_{t_d-t_n}))+o(t_d-t_n)\label{majou*}
\end{align}
where we used the monotonicity of $c$ on $(t_d-\varepsilon,t_d)$ for the first inequality and 
a reasoning similar to the one made when $D$ is concave in the proof of Lemma \ref{limnul} for the last equality.

Assume that $D$ is not the identity function which implies $x^\star<+\infty$. Using the monotonicity of both $g$ and $\frac{D(x)}{x}$, one gets that for $(t,x)\in [0,t_d)\times\R_+^*$, $\E(e^{-r{(t_d-t)}}g(\S^x_{t_d-t}))$ is not greater than
\begin{align*}
   &\E\left(e^{-r(t_d-t)}(K-\S^x_{t_d-t}+D(\S^x_{t_d-t}))1_{\{\S^x_{t_d-t}\leq x^\star\}}\right)+e^{-r(t_d-t)}g(x^\star)\P(\S^x_{t_d-t}>x^\star)\\
&\leq e^{-r(t_d-t)}K-x+\E\left(e^{-r(t_d-t)}\S^x_{t_d-t}1_{\{\S^x_{t_d-t}>x^\star\}}\right)+\frac{x}{\mu}+e^{-r(t_d-t)}g(x^\star)\P(\S^x_{t_d-t}>x^\star).
\end{align*}
Hence for $x\in (0,x^\star/2)$, $\E(e^{-r{(t_d-t)}}g(\S^x_{t_d-t}))\leq K-x+\frac{x}{\mu}-rK(t_d-t)+o(t_d-t)$ with the $o(t_d-t)$ not depending on $x\leq x^\star/2$. This inequality still holds when $D$ is the identity function, since then $\mu=1$ and $\E(e^{-r{(t_d-t)}}g(\S^x_{t_d-t}))=e^{-r(t_d-t)}K$.

With \eqref{majou*}, one deduces that
$u(t_n,x_n)\leq K-x_n+rK\frac{\tilde{\mu}-\mu}{2\mu}(t_d-t_n)+o(t_d-t_n)$. Hence for $n$ large enough $u(t_n,x_n)<K-x_n$ which provides the desired contradiction.\end{adem}

\begin{adem}[of Proposition \ref{propcle}]Let $0\leq t\leq s<t_d$, $x>0$
  and $\tau\in{\cal T}_{[0,t_d-t]}$ be such that
   $u(t,x)=\E\left(e^{-r\tau}(K-\S^x_\tau)^+1_{\{\tau<t_d-t\}}+e^{-r(t_d-t)}g(\S^x_{t_d-t})1_{\{\tau=t_d-t\}}\right)$.
Since by Lemma \ref{propg}, $\forall x>0$, $g(x)\geq(K-x)^+$,
\begin{align}
   u(t,x)&\leq
\E\left(e^{-r\tau}(K-\S^x_\tau)^+1_{\{\tau<t_d-s\}}+e^{-r\tau}g(\S^x_{\tau})1_{\{\tau\geq
    t_d-s\}}\right)\notag\\
&=\E\left(e^{-r\tau}(K-\S^x_\tau)^+1_{\{\tau<t_d-s\}}+e^{-r(t_d-s)}g(\S^x_{t_d-s})1_{\{\tau\geq
    t_d-s\}}\right)\notag\\&+\E\left(1_{\{\tau>
    t_d-s\}}\left(e^{-r\tau}g(\S^x_{\tau})-e^{-r(t_d-s)}g(\S^x_{t_d-s})\right)\right)\label{crois}.\end{align}
By Tanaka's formula,
$$d(\S^x_v-D(\S^x_v))=(1-D'_-(\S^x_v))d\S^x_v-\frac{1}{2}\int_0^{+\infty}D''(da)
dL^a_v(\S^x).$$
In particular $d\left<\S^x-D(\S^x)\right>_v=(\sigma \S^x_v(1-D'_-(\S^x_v)))^2dv$.
The function $x\mapsto \U(t_d,x)$ is convex and $C^1$ on $[0,+\infty)$
and $C^2$ on $[0,\bar{c}(t_d))$ and $(\bar{c}(t_d),+\infty)$. Hence its second order
distribution derivative is equal to $\partial_{22}\U(t_d,x)dx$ where, by
convention, $\partial_{22}\U(t_d,\bar{c}(t_d))=0$. 
Applying again Tanaka's formula and the occupation times formula, one
deduces that
$$dg(\S^x_v)=\partial_2\U(t_d,\S^x_v-D(\S^x_v))d(\S^x_v-D(\S^x_v))+\frac{\sigma^2}{2}\partial_{22}\U(t_d,\S^x_v-D(\S^x_v))((1-D'_-(\S^x_v))\S^x_v)^2dv.$$
One deduces that for $\gamma$
defined in Lemma \ref{propg},
\begin{align}
   d(e^{-rv}g(\S^x_v))=e^{-rv}\bigg(&
  \partial_2\U(t_d,\S^x_v-D(\S^x_v))\left[(1-D'_{-}(\S^x_v))\sigma \S^x_vdW_v-\frac{1}{2}\int_0^{+\infty}D''(da)
dL^a_v(\S^x)\right]\notag\\&+\gamma(\S^x_v)dv\bigg).\label{dgactu}
\end{align} The process $(\int_0^ve^{-rw}\sigma \S^x_w
\partial_2\U(t_d,\S^x_w-D(\S^x_w))(1-D'_{-}(\S^x_w))dW_w)_v$
is a martingale since $\partial_2 \U\in[-1,0]$ by \eqref{lip} and
$(1-D'_-)\in[0,1]$ according to Lemma \ref{propg}. With \eqref{crois}, one deduces that
\begin{align}
 u(s,x)-u(t,x)\geq -\E\left(1_{\{\tau> t_d-s\}}\int_{t_d-s}^{\tau}e^{-rv}\gamma(\S^x_v)dv\right)=-\E\left(\int_{t_d-s}^{t_d-t}1_{\{\tau>v\}}e^{-rv}\gamma(\S^x_v)dv\right)\label{ust}.\end{align}
One easily deduces \eqref{minodertemps} and, since by Lemma
\ref{propg}, $C\stackrel{\rm def}{=}\sup_{x>0}\gamma(x)<+\infty$ and $\gamma(x)$ is not
greater than
$-rK$ for $x<x^\star$, 
\begin{align}
 u(s,x)\geq u(t,x)+\int_{t_d-s}^{t_d-t}e^{-rv}\left(rK\P(\tau>v,\S^x_v<x^\star)-C\P(\tau>v,\S^x_v\geq
x^\star)\right)dv.\label{croisu}\end{align}

Define $\hat{c}(s)=\sup_{v\in[t_d-s,t_d)}c(v)$ and 
let $\alpha\in (0,t_d]$ be such that 
$\hat{c}(\alpha)<x^\star$. The existence of
$\alpha$ is ensured
by Lemma \ref{limnul} which applies since, according to the proof of
Lemma \ref{propg}, the function $D$ is continuous and both $D$ and
$x-D(x)$ are non-decreasing. We now choose $t\in
[t_d-\alpha,t_d)$ and $x\in (c(t),y)$ where $y\in(\hat{c}(\alpha),x^\star)$. One has
$\tau=\inf\{v\in[0,t_d-t):\S^x_v\leq c(t+v)\}$ with
convention $\inf\emptyset=t_d-t$. Let $\tau_y=\inf\{v\geq 0:\S^x_v=y\}$. For $v\in[0,t_d-t)$, by the Markov
property, one has
\begin{align*}
   \P(\tau>v,\S^x_v\geq x^\star)=\P(\tau>v,\tau_y\leq v,\S^x_v\geq
   x^\star)\leq \P(\tau_y\leq v,\tau>\tau_y)\P\left(\max_{w\in[0,v]}\S^1_w\geq x^\star/y\right).
\end{align*}
In the same time,
\begin{align*}
   \P(\tau>v)\geq \P(\tau_y\leq v,\tau>v)\geq \P(\tau_y\leq
   v,\tau>\tau_y)\P\left(\min_{w\in[0,v]}\S^1_w>\hat{c}(\alpha)/y\right).
\end{align*}
Combining both inequalities, one obtains
$$\P(\tau>v,\S^x_v\geq x^\star)\leq
\P(\tau>v)\frac{\P\left(\max_{w\in[0,\alpha]}\S^1_w\geq
    x^\star/y\right)}{\P\left(\min_{w\in[0,\alpha]}\S^1_w>\hat{c}(\alpha)/y\right)}.$$
 The ratio $\frac{\P\left(\max_{w\in[0,\beta]}\S^1_w\geq
    z\right)}{\P\left(\min_{w\in[0,\beta]}\S^1_w>\eta\right)}$ equals
$$\frac{N((\frac{r}{\sigma}-\frac{\sigma}{2})\beta-\frac{\log z}{\sigma})+e^{\frac{2\log z}{\sigma}(\frac{r}{\sigma}-\frac{\sigma}{2})}N(-(\frac{r}{\sigma}-\frac{\sigma}{2})\beta-\frac{\log z}{\sigma})}{1-N(\frac{\log \eta}{\sigma}-(\frac{r}{\sigma}-\frac{\sigma}{2})\beta)-e^{\frac{2\log \eta}{\sigma}(\frac{r}{\sigma}-\frac{\sigma}{2})}N(\frac{\log \eta}{\sigma}+(\frac{r}{\sigma}-\frac{\sigma}{2})\beta)}$$
and for $\beta>0$ and $z>1>\eta>0$
this converges to $0$ as $\beta$ and $\eta$ go to $0^+$ while $z$ goes to $+\infty$.
Since by Lemma \ref{limnul}, $\hat{c}(\alpha)$ converges to $0$ as $\alpha$
goes to $0^+$, one may choose positive constants $y,\alpha$ such that
$y\in(\hat{c}(\alpha),x^\star)$ and
$$\frac{\P\left(\max_{w\in[0,\alpha]}\S^1_w\geq
    x^\star/y\right)}{\P\left(\min_{w\in[0,\alpha]}\S^1_w>\hat{c}(\alpha)/y\right)}\leq \frac{rK}{rK+C}.$$
With $\P(\tau>v,\S^x_v<x^\star)=\P(\tau>v)-\P(\tau>v,\S^x_v\geq x^\star)$
and \eqref{croisu}, we conclude that 
$$\forall t_d-\alpha\leq t\leq s<t_d,\;\forall x\in (0,y),\;u(t,x)\leq
u(s,x).$$
Since for $t\in (0,t_d)$ and $x>c(t)$,
$\frac{\sigma^2x^2}{2}\partial_{xx}u(t,x)=-\partial_t
u(t,x)-rx\partial_xu(t,x)+ru(t,x)$ with $\partial_xu\in [-1,0]$
according to \eqref{lip} and $u\leq
K$, \eqref{majodersec} easily follows from \eqref{minodertemps}.
Let $t\in [0,t_d)$ be such that $c(t)>0$. For $z\geq x>c(t)$, one has
$\partial_xu(t,x)=\partial_xu(t,z)-\int_x^z\partial_{xx}u(t,y)dy$. By \eqref{lip},
$\partial_x u(t,x)\in [-1,0]$. With \eqref{majodersec}, one deduces that
$y\mapsto\partial_{xx}u(t,y)$ is integrable on
$[c(t),z]$ and the right-hand limit $\partial_x u(t,c(t)^+)$ makes
sense.\end{adem}
\begin{arem}
   When $T=+\infty$ i.e. when the Put option is perpetual, 
\begin{equation*}
   u(t_d,x)=\begin{cases}
      K-x\mbox{ if }x<\bar{c}(t_d)=\frac{-K\alpha}{1-\alpha}\\
(K-\bar{c}(t_d))(x/\bar{c}(t_d))^{\alpha}\mbox{ otherwise}
   \end{cases},\mbox{ where }\alpha=-\frac{2r}{\sigma^2}.
\end{equation*}

In the
{\sl proportional} dividend case, $\gamma(x)=-rK1_{\{x<\bar{c}(t_d)/\rho\}}$ since ${\cal A}f(x)=0$ for $f(x)=x^\alpha$.
With \eqref{ust}, one deduces that for any $x>0$, $t\mapsto u(t,x)$ is
non-decreasing on $[0,t_d)$.

In the {\sl constant} dividend case, 
\begin{equation*}
   \gamma(x)=\begin{cases}-rK\mbox{ if }x\in(0,D)\\
-r(K+D)\mbox{ if }x\in(D,\bar{c}(t_d)+D)\\-\alpha(K-\bar{c}(t_d))\bar{c}(t_d)^{-\alpha} D(rx+\frac{\sigma^2}{2}(2x-D))(x-D)^{\alpha-2}\mbox{ if }x>\bar{c}(t_d)+D
\end{cases}
\end{equation*}
is positive on $(\bar{c}(t_d)+D,+\infty)$.
\end{arem}
\section{Continuity of the exercise boundary and high contact principle}

We can now state our main result concerning the
continuity of the exercise
boundary $c(t)$ for the single dividend case. Note that it applies to the
{\sl proportional}, the {\sl constant} and the more general {\sl mixed} dividend cases.

\begin{aprop}\label{contfront}
   Assume that $D$ is a positive concave function such that $x-D(x)$ is
   non-negative. Then for $t\in[0,t_d)$ such that $c$ is
   right-continuous at $t$, the smooth contact
   property holds $\partial_x
   u(t,c(t)^+)=-1$ and $\lim_{s\to t^+}\partial_x
   u(s,c(s)^+)=-1$.\\
   If $g$ is convex, then $t\mapsto c(t)$ is right-continuous on
   $[0,t_d)$. More generally, if $D$ is such that 
\begin{align}\label{divprop0}
   \exists x_0>0,\exists\rho\in [0,1),\;\forall x\in (0,x_0),\;D(x)=(1-\rho)x,
\end{align}
then there exists an $\varepsilon\in (0,t_d]$ such that $t\mapsto c(t)$ is continuous on $(t_d-\varepsilon,t_d)$.
\end{aprop}
\begin{arem}On any open interval on which $c$ is non-decreasing, it is right-continuous by upper-semicontinuity and therefore the smooth contact
   property holds.\end{arem}
In order to prove the Proposition, we will need the following
estimations of the first order time
derivative and the second order spatial derivative of the pricing function
$u$ in the continuation region.\begin{alem}\label{majodert}
      Assume that $D$ is a non-negative concave function such that $x-D(x)$ is
   non-negative.  Then 
\begin{align}
 \forall t\in [0,t_d),\;\forall x>c(t),\;&\partial_t u(t,x)\leq
 -e^{-r(t_d-t)}\inf_{y>0}\gamma(y)+\frac{\sigma x}{2\sqrt{2\pi(t_d-t)}}\label{majoderttt}\\
  \mbox{ and }&\frac{\sigma^2x^2}{2}\partial_{xx}u(t,x)\geq e^{-r(t_d-t)}\inf_{y>0}\gamma(y)-\frac{\sigma x}{2\sqrt{2\pi(t_d-t)}}+r(K-x)^+.\label{minodersecc}
\end{align}

If $g$ is
   convex, then for $t\in [0,t_d)$, $x\mapsto u(t,x)$ is convex and for $x>c(t)$, $\partial_t u(t,x)\leq rK
   e^{-r(t_d-t)}$ and $\partial_{xx}u(t,x)\geq 0$.\\
   More generally,  under \eqref{divprop0}, there
   exists $\varepsilon\in (0,t_d]$ such that
   for all $t\in (t_d-\varepsilon,t_d)$ and for all $x\in
(c(t),c(t)+\varepsilon)$ we have $\partial_t u(t,x)\leq
rK\frac{1+e^{-r(t_d-t)}}{2}$.
 \end{alem} \begin{adem}[of Proposition \ref{contfront}]
For $t\in[0,t_d)$, $c(t)>0$ by Corollary \ref{frontdeccag}, and by
  Proposition \ref{propcle}, the following Taylor expansion makes sense
\begin{align}
\forall x\geq
c(t),\;u(t,x)=(K-c(t))+(x-c(t))\partial_xu(t,c(t)^+)+\int_{c(t)}^x(x-y)\partial_{xx}u(t,y)dy.\label{tayl}\end{align}
Substituting $z$ for $x$ in \eqref{tayl} and subtracting the result from \eqref{tayl} itself gives for $x>z\geq c(t)$
\begin{align}
\partial_xu(t,c(t)^+)=\frac{u(t,x)-u(t,z)}{x-z}-\int_{c(t)}^{z}\partial_{xx}u(t,y)dy-\frac{1}{x-z}\int_{z}^{x}(x-y)\partial_{xx}u(t,y)dy.
\label{uxct}\end{align}
If $s\in [0,t_d)$ is such that $c(s)\geq c(t)$, choosing $z=c(s)$ and computing $\partial_x u(s,c(s)^+)$ from
\eqref{uxct} written with $s$ replacing $t$, one deduces that for $x>c(s)$,
\begin{align}
   \partial_xu(s,c(s)^+)-\partial_xu(t,c(t)^+)&=\frac{1}{x-c(s)}\bigg(u(s,x)-u(t,x)+u(t,c(s))-u(s,c(s))\bigg)\notag\\&+\frac{1}{x-c(s)}\int_{c(s)}^x(x-y)(\partial_{xx}u(t,y)-\partial_{xx}u(s,y))dy\notag\\&+\int_{c(t)}^{c(s)}\partial_{xx}u(t,y)dy.\label{evolder1}
\end{align}
We decompose the proof in three steps using the above expansions. First we check that when $t_0\in [0,t_d)$ is such that $c$ is right-continuous at $t_0$, then $\lim_{t\to t_0^+}\partial_xu(t,c(t)^+)=\partial_xu(t_0,c(t_0)^+)$. In the second step, we check that when $c$ is right-continuous at $t_0$, then the smooth contact property holds at $t_0$. In the last step, we prove that $c$ is right-continuous at $t_0$ for $t_0$ close to $t_d$ under \eqref{divprop0} and with no restriction in the convex case.

{\bf Step 1 :} Let $t_0\in [0,t_d)$ be such that $c$ is right-continuous at $t_0$ and $x>c(t_0)$. For $t>t_0$ such that $c(t)<x$, by \eqref{evolder1}, $|\partial_xu(t_0,c(t_0)^+)-\partial_x u(t,c(t)^+)|$ is smaller than
\begin{align*}
   &\frac{1}{x-c(t)\vee c(t_0)}|u(t_0,x)-u(t,x)+u(t,c(t)\vee c(t_0))-u(t_0,c(t)\vee c(t_0))|\\&+\frac{1}{x-c(t)\vee c(t_0)}\int_{c(t)\vee c(t_0)}^x(x-y)|\partial_{xx}u(t,y)-\partial_{xx}u(t_0,y)|dy\\&
+\int_{c(t)\wedge c(t_0)}^{c(t)\vee c(t_0)}|\partial_{xx}u(t,y)|+|\partial_{xx}u(t_0,y)|dy
\end{align*}
By continuity of $u$, the first term converges to $0$
as $t\to t_0^+$. Moreover, \eqref{majodersec} and \eqref{minodersecc}
ensure that the second term is
arbitrarily small uniformly for $t<(t_0+t_{d})/2$ when $x$ is close
enough to $c(t_0)$. Last, with the right-continuity of $c$ at $t_0$, the third
term converges to $0$ as $t\to t_0^+$, which ensures the desired right-continuity property.\\
{\bf Step 2 :} Let us now assume that for $t_0\in[0,t_d)$ such that $c$ is right-continuous at $t_0$, $\partial_xu(t_0,c(t_0)^+)>-1$ and obtain a
contradiction. 
Let $t\in (t_0,\frac{t_0+t_d}{2})$ be such that $c(t)\leq c(t_0)$. According to \eqref{majoderttt} and \eqref{minodersecc}, there exists a constant
$C\in (0,+\infty)$ such that $u(t,c(t_0))\leq K-c(t_0)+C(t-t_0)$ and
$\int_{c(t)}^{c(t_0)}(c(t_0)-y)\partial_{xx}u(t,y)dy\geq
-C\frac{(c(t_0)-c(t))^2}{c(t)^2}$. Writing \eqref{tayl} for $x=c(t_0)$, one deduces that
$$\left(1+\partial_xu(t,c(t)^+)-C\frac{c(t_0)-c(t)}{c(t)^2}\right)(c(t_0)-c(t))\leq C(t-t_0).$$
Since $\partial_xu(t,c(t)^+)$ tends to $\partial_xu(t_0,c(t_0)^+)>-1$ as
$t\to t_0^+$ and $c$ is right-continuous at $t_0$, one deduces the existence of $\varepsilon\in (0,t_d-t_0)$
such that
\begin{align}
\forall t\in [t_0,t_0+\varepsilon],\;c(t)-c(t_0)\geq -\frac{2C(t-t_0)}{1+\partial_xu(t_0,c(t_0)^+)}.\label{minofront}
\end{align}
For $x>c(t_0)$, let $\tau_x=\inf\{s>0:\S^x_s\leq c(t_0+s)\}\wedge (t_d-t_0)$
denote the stopping time such that 
$$u(t_0,x)=\E\left(e^{-r\tau_x}(K-\S^x_{\tau_x})^+1_{\{\tau_x<t_d-t_0\}}+e^{-r(t_d-t_0)}g(\S^x_{\tau_x})1_{\{\tau_x=t_d-t_0\}}\right).$$
One has
$u(t_0,c(t_0))\geq\E\left(e^{-r\tau_x}(K-\S^{c(t_0)}_{\tau_x})^+1_{\{\tau_x<t_d-t_0\}}+e^{-r(t_d-t_0)}g(\S^{c(t_0)}_{\tau_x})1_{\{\tau_x=t_d-t_0\}}\right)$.
Computing the difference, using the monotonicity of $g$ and the
Lipschitz continuity of $y\mapsto (K-y)^+$ one deduces that
\begin{align}
   \frac{u(t_0,x)-u(t_0,c(t_0))}{x-c(t_0)}\leq -\E\left(e^{-r\tau_x}\S^{1}_{\tau_x}1_{\{\tau_x<t_d-t_0\}}\right).\label{majax}
\end{align}
By \eqref{minofront}, $\tau_x\leq \tilde{\tau}_x\stackrel{\rm
  def}{=}\inf\{s\in(0,\varepsilon]:\S^x_s\leq
c(t_0)-2Cs/(1+\partial_xu(t_0,c(t_0)^+))\}\wedge (t_d-t_0)$. When $x$
tends to $c(t_0)^+$, $\tilde{\tau}_x$ converges a.s. to $\inf\{s\in(0,\varepsilon]:\S^1_s<
1-2Cs/(c(t_0)(1+\partial_xu(t_0,c(t_0)^+)))\}\wedge (t_d-t_0)$ which is
equal to $0$ according to the iterated
logarithm law satisfied by the Brownian motion $W$. Hence $\tau_x$ converge a.s. to $0$ as $x\to c(t_0)^+$. Since
$\E(\sup_{s\in[0,t_d-t_0]}\S^1_s)<+\infty$, by
Lebesgue's theorem, the
right-hand-side of \eqref{majax} converges to $-1$ as $x\to c(t_0)^+$
which implies the desired contradiction : $\partial_x
u(t_0,c(t_0)^+)\leq -1$.\par

{\bf Step 3 :} Let $t_0\in [0, t_d)$
be such that $c$ is not right-continuous at $t_0$. We are going to derive a contradiction when $g$ is convex or $t_0$ close to $t_d$ under \eqref{divprop0}. The continuity of $c$ on a left-hand neighbourhood of $t_d$ then follows from the left-continuity stated in Corollary \ref{frontdeccag}. By the upper-semicontinuity of $c$ and the positivity of $\inf_{t\in[0,\frac{t_0+t_d}{2}]}c(t)$ stated in Corollary \ref{frontdeccag}, there exists a sequence $(s_k)_{k\in\N}$ in $(t_0,t_d)$ converging to $t_0$ as $k\to\infty$ and such that $\lim_{k\to\infty}c(s_k)\in (0,c(t_0))$. 
Let $x,z\in(\lim_{k\to\infty}c(s_k),c(t_0))$ with $x>z$.
For $k$ large enough $c(s_k)<z$ and we may use \eqref{uxct} for $t=s_k$. The left-hand-side is not smaller than $-1$. When $k$ tends to $\infty$, by
continuity of $u$, the first term in the right-hand-side tends to
$\frac{K-x-(K-z)}{x-z}=-1$. Moreover by \eqref{minodersecc}, there is a constant $C\in (0,+\infty)$ not depending on $k$ such that
\begin{align*}
   \int_{c(s_k)}^z\partial_{xx}u(s_k,y)dy+\frac{1}{x-z}\int_{z}^{x}(x-y)\partial_{xx}u(s_k,y)dy\geq -\frac{C}{c^2(s_k)}\left(2(z-c(s_k))+(x-z)\right).
\end{align*}
Hence $\limsup_{k\to\infty}\partial_xu(s_k,c(s_k)^+)\leq -1+\frac{C}{\lim_{k\to \infty}c^2(s_k)}\left(x+z-2\lim_{k\to \infty}c(s_k)\right)$ and one deduces \begin{equation}
   \lim_{k\to\infty}\partial_xu(s_k,c(s_k)^+)=-1\label{limkderfront}
\end{equation} by letting $x$ and $z$ go to $\lim_{k\to \infty}c(s_k)$.
By \eqref{tayl} and Proposition \ref{propcle},
\begin{align*}
   \forall
x>c(t),\;\int_{c(t)}^x y\partial_{xx}u(t,y)dy&=
   x\partial_xu(t,x)-c(t)\partial_xu(t,c(t)^+)-u(t,x)+u(t,c(t))\\&=x\partial_xu(t,x)-u(t,x)+K-c(t)\left(1+\partial_xu(t,c(t)^+)\right).
\end{align*}
With the equality $\partial_t u(t,x)+{\cal A}u(t,x)=0$ and 
Lemma \ref{majodert}, one deduces that for $t$ and $t_0$ close to $t_d$ under \eqref{divprop0}
and with no restriction in the convex case,
\begin{align}
\forall
x\in(c(t),c(t_0)),\;\frac{\sigma^2x^2}{2}\partial_{xx}u(t,x)&+r\int_{c(t)}^x
y\partial_{xx}u(t,y)dy=rK-\partial_t
u(t,x)-rc(t)(1+\partial_{x}u(t,c(t)^+)\notag\\&\geq
\frac{rK(1-e^{-r(t_d-t)})}{2}-rc(t)\left(1+\partial_xu(t,c(t)^+)\right).\label{minodersec}\end{align}
According to \eqref{majodersec}, there is a finite constant $C$ such
that $\forall t\in[0,t_d),\;\forall
x\in(c(t),c(t_0)]$, $\partial_{xx}u(t,y)\leq \frac{C}{y}$ so 
$r\int_{c(t)}^x y\partial_{xx}u(t,y)dy\leq rK(1-e^{-r(t_d-t)})/8$ if we
take $x\leq c(t_0)\wedge c(t)e^{\frac{K(1-e^{-r(t_d-t)})}{8C}}$.
With
\eqref{limkderfront} and \eqref{minodersec}, one deduces that for $t_0$ close to $t_d$ under \eqref{divprop0}
and with no restriction in the convex case, for $k$ large enough,
\begin{itemize}
   \item  $\forall y\in\left(c(s_k),c(t_0)\wedge c(s_k)e^{\frac{K(1-e^{-r(t_d-s_k)})}{8C}}\right),$ $\frac{\sigma^2y^2}{2}\partial_{xx}u(s_k,y)\geq \frac{rK(1-e^{-r(t_d-s_k)})}{4}$,
\item and therefore for $x,z\in\left(\lim_{l\to \infty}c(s_l),c(t_0)\wedge \lim_{l\to \infty}c(s_l)e^{\frac{K(1-e^{-r(t_d-t_0)})}{16C}}\right)$ with $x>z$,
\begin{align*}
  \int_{c(s_k)}^z\partial_{xx}u(s_k,y)dy+\frac{1}{x-z}\int_{z}^{x}(x-y)\partial_{xx}u(s_k,y)dy\geq
\frac{rK(1-e^{-r(t_d-s_k)})}{4\sigma^2x^2}(x+z-2c(s_k)).
\end{align*} \end{itemize}

Taking the limit $k\to\infty$ in \eqref{uxct} written for $t=s_k$, we now obtain $\limsup_{k\to\infty}\partial_x
u(s_k,c(s_k)^+)<-1$, which contradicts \eqref{limkderfront}.\end{adem}

\begin{adem}[of Lemma \ref{majodert}]
Let $t\in [0,t_d)$. When $g$ is convex, since $x\mapsto (K-x)^+$ is also convex, for each
stopping time $\tau\in{\cal T}_{[0,t_d-t]}$, $x\mapsto
\E(e^{-r\tau}(K-\S^x_{\tau})^+1_{\{\tau<t_d-t\}}+e^{-r(t_d-t)}g(\S^x_{t_d-t})1_{\{\tau=t_d-t\}})$
is convex. So $x\mapsto u(t,x)$ which is equal to the supremum over
$\tau$ of the previous functions is convex.\\
   Let now $0\leq t\leq s<t_d$, $x>0$ and $\tau\in{\cal
     T}_{[0,t_d-s]}$ be such that
   $$u(s,x)=\E\left(e^{-r\tau}(K-\S^x_\tau)^+1_{\{\tau<t_d-s\}}+e^{-r(t_d-s)}g(\S^x_{t_d-s})1_{\{\tau=t_d-s\}}\right).$$Since $u(t,x)\geq
\E\left(e^{-r\tau}(K-\S^x_\tau)^+1_{\{\tau<t_d-s\}}+e^{-r(t_d-t)}g(\S^x_{t_d-t})1_{\{\tau=
    t_d-s\}}\right)$, one has
\begin{align*}
   u(t,x)-u(s,x)&\geq
\E\left(1_{\{\tau=t_d-s\}}\left(e^{-r(t_d-t)}g(\S^x_{t_d-t})-e^{-r(t_d-s)}g(\S^x_{t_d-s})\right)\right).
\end{align*}
When $g$ is convex, according to Lemma \ref{propg}, ${\mathcal A}g$ is
a function bounded from below by $-rK$, the right-hand-side is equal
to $\E\left(1_{\{\tau=t_d-s\}}\int_{t_d-s}^{t_d-t}e^{-rv}{\cal
    A}g(\S^x_v)dv\right)$, so one easily concludes.
In general, by \eqref{dgactu} and the martingale property of the process $(\int_0^ve^{-rw}\sigma \S^x_w
\partial_2\U(t_d,\S^x_w-D(\S^x_w))(1-D'_{-}(\S^x_w))dW_w)_v$, the
previous inequality writes 
\begin{align}
   u(t,x)&-u(s,x)\notag\\&\geq \E\left(1_{\{\tau=t_d-s\}}\int_{t_d-s}^{t_d-t}e^{-rv}\left[\gamma(\S^x_v)dv-\frac{\partial_2\U(t_d,\S^x_v-D(\S^x_v))}{2}\int_0^\infty
  D''(da)dL^a_v(\S^x)\right]\right).\label{ut-us}
\end{align}
Since $\partial_2\U(t_d,y)\geq -1$, using
  the occupation times formula, one deduces that
$$u(s,x)-u(t,x)\leq\int_{t_d-s}^{t_d-t}e^{-rv}\left(-\inf_{y>0}\gamma(y)-\int_{0}^{+\infty}
  \frac{\sigma^2 a^2}{2}p(v,a)D''(da)\right)dv.$$
Since $D(x)$ and $x-D(x)$ are both non-decreasing, $D''((0,+\infty))\geq
-1$. Using moreover
$$\forall v\in [0,t_d-t],\;\forall a>0,\;a^2p(v,a)=\frac{xe^{rv}}{\sigma\sqrt{2\pi
    v}}e^{-\frac{(\log(a/x)-(r+\frac{\sigma^2}{2})v)^2}{2\sigma^2
    v}}\leq\frac{xe^{rv}}{\sigma\sqrt{2\pi
    v}},$$
one deduces \eqref{majoderttt}. The inequality \eqref{minodersecc} follows since for $x>c(t)$ we have $\frac{\sigma^2x^2}{2}\partial_{xx}u(t,x)=-\partial_t
u(t,x)-rx\partial_xu(t,x)+ru(t,x)\geq -\partial_t
u(t,x)+r(K-x)^+$.

Assume  \eqref{divprop0}. Then $\gamma$ is equal to $-rK$ on
$(0,x_0\wedge x^\star)$, $D''((0,x_0))=0$ and \eqref{ut-us} implies that
$$u(s,x)-u(t,x)\leq\int_{t_d-s}^{t_d-t}e^{-rv}\left(rK-(\inf_{y>0}\gamma(y)+rK)\P(\S^x_v\geq
  x_0\wedge x^\star)-\int_{x_0}^{+\infty} \frac{\sigma^2 a^2}{2}p(v,a)D''(da)\right)dv.$$
For $x\in (0,x_0e^{-(r+\frac{\sigma^2}{2})(t_d-t)}]$, one has $\forall v\in [0,t_d-t],\;\forall a\geq x_0,\;a^2p(v,a)\leq \frac{xe^{rv}}{\sigma\sqrt{2\pi
    v}}e^{-\frac{(\log(x_0/x)-(r+\frac{\sigma^2}{2})v)^2}{2\sigma^2
    v}}$. For $t$ close enough to $t_d$ we have that
$c(t)<x_0e^{-(r+\frac{\sigma^2}{2})(t_d-t)}$ by Lemma \ref{limnul}  and
for $x\in
(c(t),x_0e^{-(r+\frac{\sigma^2}{2})(t_d-t)})$,
\begin{align*}
   \partial_t u(t,x)\leq&
e^{-r(t_d-t)}\left(rK-(\inf_{y>0}\gamma(y)+rK)N\left(\frac{\log(x/(x_0\wedge
      x^\star))+(r-\frac{\sigma^2}{2})(t_d-t)}{\sigma\sqrt{t_d-t}}\right)\right)
  \\&+\frac{\sigma x}{2\sqrt{2\pi
    (t_d-t)}}e^{-\frac{(\log(x_0/x)-(r+\frac{\sigma^2}{2})(t_d-t))^2}{2\sigma^2
    (t_d-t)}}.
\end{align*}
Bounding from above the two last terms like in the derivation of the upper-bound for $c(t)$ in the
proof of Lemma \ref{limnul}, one deduces the last assertion.
\end{adem}

\section{Conclusions and Further Research}
 
 We have proven local results concerning the regularity of the exercise
boundary for a dividend-paying asset. Even in the simplest case of
{\sl proportional dividends}, it would be of great interest to
prove the following feature observed in numerical calculations: for a single dividend payment, when
$t_d$ is large, the exercise boundary is non-decreasing for small times
and monotonicity seems to change only once before $t_d$. We also would like to extend the results that we have obtained for multiple dividend payments in the {\sl proportional} case to more general functions $D^i$. The key issue in this perspective is to derive global estimates on the derivatives of the value function $u^1$ before $t_d^1$ to replace those which follow from the convexity in the variable $x$ in the {\sl proportional} case. 

 Another interesting matter to investigate would be the optimal exercise
boundary for the alternative model for dividends known as the Escrowed Model. 
As we have shown in Remark \ref{Escrowed}, this boundary is zero on an interval
with strictly positive length before every dividend date, but other properties
of this boundary have yet to be established.
 

\begin{appendix}
\section{Proof of Theorem \ref{multiple}}
The two first statements can easily be deduced by respectively adapting the comparison argument given at the beginning of the proof of Corollary \ref{frontdeccag} and the proof of Lemma \ref{limnul}.\\

Let us now consider the case of multiple proportional dividends. We will prove by induction on $i$ that the statement holds together with the following lemma.
\begin{alem}
   If for all $1\leq j\leq i$ we have $D^j(x)=(1-\rho_j)x$ for some $\rho_j\in (0,1)$ then $g^i$ is convex and $C^1$ on $\R_+$ and $C^2$ on $[0,x^\star_{i})\cup(x^\star_{i},+\infty)$ for $x^\star_{i}\stackrel{\rm def}{=}\frac{c^{i-1}(t_d^{i})}{\rho_{i}}$. Moreover, the function $\gamma_{i}(x)\stackrel{\rm def}{=}{\cal A}g^i(x)$ is equal to $-rK$ on $[0,x_{i}^\star)$, not smaller than $-rK$ and bounded on $(x^\star_{i},+\infty)$ and satisfies \begin{align}
   \forall t\in[0,t_d^{i}),\;\forall x>c^{i}(t),&\;-e^{-r(t_d^{i}-t)}\sup_{y>0}\gamma^+_{i}(y)\leq\partial_tu^{i}(t,x)\leq e^{-r(t_d^{i}-t)}rK\label{contdertempsi}\\&\mbox{and }0\leq \frac{\sigma^2 x^2}{2}\partial_{xx}u^{i}(t,x)\leq e^{-r(t_d^{i}-t)}\sup_{y>0}\gamma^+_{i}(y)+rK.\label{estimgami}
\end{align}
\end{alem}
For $i=1$, the result is a consequence of Propositions \ref{propcle} and \ref{contfront}, Corollary \ref{frontdeccag} and Lemma \ref{majodert}, the refinement over \eqref{majodersec} in the last inequality in \eqref{estimgami} following from the monotonicity of $x\mapsto x\partial_xu^1(t,x)-u^1(t,x)$ which is a consequence of the convexity of $x\mapsto u^1(t,x)$.

Assume the induction hypothesis to be true for a certain $i\geq 1$. Then $x\mapsto g^{i+1}(x)=u^i(t^{i+1}_d,\rho_{i+1}x)$ is convex and arguing like in the beginning of the proof of Lemma \ref{majodert}, one obtains that for $t\in[0,t_d^{i+1})$, $x\mapsto u^{i+1}(t,x)=\sup_{\tau \in {\cal T}_{[0,t_d^{i+1}-t]}} {\Bbb E}[ e^{-r\tau}(\, (K-\S_{\tau}^x)^+1_{\{\tau<t^{i+1}_d-t\}}+g^{i+1}(\S_{t^{i+1}_d-t}^x)1_{\{\tau=t_d^{i+1}-t\}}  \, ) ]$ is convex and nonincreasing.
The function $g^{i+1}$ is $C^1$ on $\R_+$ by the smooth contact property for $u^i$ at time $t^{i+1}_d$ and $C^2$ on $[0,x^\star_{i+1})\cup(x^\star_{i+1},+\infty)$ by the regularity properties of $u^i$ stated in Lemma \ref{operatorcontinuationregion}. Moreover the function $\gamma_{i+1}(x)={\cal A}u^i(t_d^{i+1},.)(\rho_{i+1}x)$ is equal to $-rK$ on $[0,x_{i+1}^\star)$, not smaller than $-rK$ and bounded on $(x^\star_{i+1},+\infty)$ respectively by convexity of $x\mapsto u^i(t_d^{i+1},x)$ and by the lower bound in \eqref{contdertempsi} combined with the equality $\partial_t u^i(t_d^{i+1},x)+{\cal A}u^i(t_d^{i+1},x)=0$ which is satisfied for $x>c^i(t_d^{i+1})$. One may now adapt the proofs of Proposition \ref{propcle}, Lemma \ref{majodert} and Corollary \ref{frontdeccag} to check 
that the exercise boundary $c^{i+1}(t)$ is non-increasing and equivalent to $\frac{rK(t_d^{i+1}-t)}{1-\rho_{i+1}}$ in a left-hand neighbourhood of $t_d^{i+1}$ and that \eqref{contdertempsi} and \eqref{estimgami} hold with $i+1$ replacing $i$. Next, with these bounds on the derivatives of $u^{i+1}$, one adapts the proof of Proposition \ref{contfront} to obtain right-continuity of the exercise boundary $c^{i+1}$ on $[0,t_d^{i+1})$ and smooth contact : $\forall t\in[0,t_d^{i+1})$, $\partial_xu^{i+1}(t,c^{i+1}(t)^+)=-1$. This proves the statement for $i+1$ and concludes the proof.

\end{appendix}

\end{document}